\documentclass[twocolumn,trackchanges]{aastex63}

\usepackage{natbib}
\usepackage{amsmath}
\usepackage{braket}
\usepackage{color}
\hypersetup{
 colorlinks=true,%
 linkcolor=black,
 citecolor=blue,
}

\begin{document}
\def\vec#1{\mbox{\boldmath $#1$}}
\newcommand{\average}[1]{\ensuremath{\langle#1\rangle} }

\title{\large{\bf Alfv\'en-wave driven magnetic rotator winds from low-mass stars I: \\
rotation dependences of magnetic braking and mass-loss rate}}

	\author{Munehito Shoda}
	\affiliation{National Astronomical Observatory of Japan, National Institutes of Natural Sciences, 2-21-1 Osawa, Mitaka, Tokyo, 181-8588, Japan}
	\author{Takeru K. Suzuki}
	\affiliation{School of Arts \& Science, The University of Tokyo, 3-8-1 Komaba, Meguro, Tokyo, 153-8902, Japan}
	\author{Sean P. Matt}
	\affiliation{University of Exeter, Department of Physics \& Astronomy, Physics Bldg., Stocker Road, Exeter EX4 4QL, UK}
	\author{Steven R. Cranmer}
	\affiliation{Department of Astrophysical and Planetary Sciences, Laboratory for Atmospheric and Space Physics, University of Colorado, Boulder, CO 80309, USA}
	\author{Aline A. Vidotto}
	\affiliation{School of Physics, Trinity College Dublin, the University of Dublin, Dublin-2, Ireland}
	\author{Antoine Strugarek}
	\affiliation{Department of Astrophysics \& UMR AIM, CEA Paris-Saclay, CNRS/INSU, University of Paris 7, Gif-sur-Yvette, France}
	\author{Victor See}
	\affiliation{University of Exeter, Department of Physics \& Astronomy, Physics Bldg., Stocker Road, Exeter EX4 4QL, UK}
	\author{Victor R\'eville}
	\affiliation{IRAP, Universit\'e Toulouse III - Paul Sabatier, CNRS, CNES, Toulouse, France}
	\author{Adam J. Finley}
	\affiliation{University of Exeter, Department of Physics \& Astronomy, Physics Bldg., Stocker Road, Exeter EX4 4QL, UK}
	\author{Allan Sacha Brun}
	\affiliation{Department of Astrophysics \& UMR AIM, CEA Paris-Saclay, CNRS/INSU, University of Paris 7, Gif-sur-Yvette, France}
	
	\correspondingauthor{Munehito Shoda}
	\email{munehito.shoda@nao.ac.jp}

\begin{abstract}

  Observations of stellar rotation show that low-mass stars lose angular momentum during the main sequence.
  We simulate the winds of Sun-like stars with a range of rotation rates, covering the fast and slow magneto-rotator regimes, including the transition between the two. 
  We generalize an Alfv\'en-wave driven solar wind model that builds on previous works by including the magneto-centrifugal force explicitly.
  In this model, the surface-averaged open magnetic flux is assumed to scale as $B_\ast f^{\rm open}_\ast \propto {\rm Ro}^{-1.2}$, where $f^{\rm open}_\ast$ and ${\rm Ro}$ are the surface open-flux filling factor and Rossby number, respectively.
  We find that, 
  1. the angular momentum loss rate (torque) of the wind is described as $\tau_w \approx 2.59 \times 10^{30} {\rm \ erg} \ \left( \Omega_\ast / \Omega_\odot \right)^{2.82}$, yielding a spin-down law $\Omega_\ast \propto t^{-0.55}$.
  2. the mass-loss rate saturates at $\dot{M}_w \sim 3.4 \times 10^{-14} M_\odot {\rm \ yr^{-1}}$, due to the strong reflection and dissipation of Alfv\'en waves in the chromosphere.
  This indicates that the chromosphere has a strong impact in connecting the stellar surface and stellar wind.
  Meanwhile, the wind ram pressure scales as $P_w \propto \Omega_\ast^{0.57}$, 
  which is able to explain the lower-envelope of the observed stellar winds by Wood et al.
  3. the location of the Alfv\'en radius is shown to scale in a way that is consistent with 1D analytic theory. 
  Additionally, the precise scaling of the Alfv\'en radius matches previous works which used thermally-driven winds. 
  Our results suggest that the Alfv\'en-wave driven magnetic rotator wind plays a dominant role in the stellar spin-down during the main-sequence.
 		
\end{abstract}


\section{Introduction} \label{sec:introduction}

The dynamo process yields stellar magnetic fields \citep{Leigh69,Brun004,Hotta16} that give rise to activity 
such as coronal heating \citep{Alfve47,Oster61,Parke88,Rappa08}, stellar winds \citep{Parke58,Velli94}, flares and coronal mass ejections \citep{Argir19,Notsu19,Toriu19}.
Stellar activity of this type is observed to decay over the lifetime of a star \citep{Skuma72,Gudel97,Gudel07,Vidot14a}.
Understanding such long-term evolution is one of the most important challenges in astronomy,
especially in the context of stellar influences on the habitability of exoplanet as the erosion of planetary atmosphere is affected by stellar activity \citep{Lamme10,Johns15b,Garra16,Johns19,Allan19,Airap20,Vidot20}.
In order to understand activity evolution, we first need to understand the evolution of stellar rotation since (differential) rotation and convection is the ultimate origin of magnetic energy in low-mass stars.

It is widely known that low-mass stars spin down over their lifetimes \citep{Schat62,Kraft67}, approximately as $\Omega \propto t^{-1/2}$ over the age range of $\sim 10^8 {\rm \ yr}$ to the age of the Sun \citep{Skuma72}.
This stellar spin-down is due to the angular-momentum loss caused by magnetized stellar winds (magnetic braking) \citep{Weber67,Sakur85,Kawal88}.
Magnetic braking governs the long-term variations in stellar rotation, and thus the stellar dynamo process,
which in turn affects the intensity and structure of the stellar wind.
In this way, the interplay between the stellar dynamo and stellar wind regulates the rotational evolution of stars \citep{Brun017}.

As well as being an indicator for dynamo efficiency, stellar rotation is an important fundamental quantity that can be used as a stellar age diagnostic.
Since magnetic braking is stronger for faster rotators,
the rotation periods of low-mass stars are often found to converge onto a sequence defined by mass and age, regardless of their initial rotation rates \citep{Irwin09}.
For example, stars with $M_\ast \gtrsim 0.5 M_\odot$ are known to have rotationally converged by the age of the Hyades cluster \citep{Radic87,Delor11}.
The use of rotation as a proxy for age in this way is known as gyrochronology \citep{Barne03,Barne07,Barne10}.
An alternative diagnostic based on magnetic field strength instead of rotation rate has also been proposed \citep{Vidot14a}.
Gyrochronology mainly appears to be applicable to middle-aged stars, i.e. $t_{\rm age} \lesssim 2.5 {\rm \ Gyr}$ for Sun-like stars \citep{Meibo15}, 
while recent asteroseismic studies suggest that the stellar age-color-rotation relation may deviate from gyrochronology for stars older than the age of the Sun \citep{Davie15,Angus15,Sader16}.
To understand what causes the break-down of gyrochronology, 
we first need to correctly model the mechanism by which stars lose angular momentum.

The evolution of stellar rotation periods is governed by several physical processes,
such as disc-locking, core-envelope decoupling, internal-structure evolution, and magnetic braking \citep{Galle13,Galle15}.
The magnetic braking plays a dominant role in the net angular-momentum loss during the main-sequence.
In contrast to observation-based approach to mass-loss \citep{Johns15a,Ahuir20} and angular momentum loss rate \citep{Matt015},
we aim to model them in a physics-based way.
One problem of physics-based stellar wind models is that 
the scaling laws of the mass-loss rate \citep{Schro05,Holzw07,Suzuk07a,Cranm11,Suzuk18} and the Alfv\'en radius \citep{Kawal88,Matt008,Matt012,Revil15,Finle17,Finle18a} 
have been discussed independently \citep[note that the torque is a function of the mass-loss rate, Alfv\'en radius and rotation rate, see][]{Weber67}.
However, both the mass-loss rate and the Alfv\'en radius vary with stellar wind density, and therefore should be modeled simultaneously.

The mass-loss rate is determined by the energy balance in the chromosphere and the corona, 
while the Alfv\'en radius is related to the large-scale magnetism of the star and stellar wind acceleration.
Thus, in order to simultaneously model the mass-loss rate and Alfv\'en radius, we need
1. to resolve the chromosphere and waves therein (typical spatial scale $\sim$ a few $100 \ {\rm km}$) and
2. a simulation domain that is sufficiently large to cover the wind acceleration (typical spatial scale $\sim$ a few $10R_\ast$ or more), which requires typically $10^{4-5}$ grid points in the radial direction.
For this reason,
we make use of a one-dimensional solar wind model that satisfies the aforementioned demand 
and generalize it to stellar wind by explicitly taking into account the rotation effect.
This model allows us to investigate the dependence of stellar wind parameters (mass-loss rate, Alfv\'en radius, torque)
on the stellar rotation rate.
A goal of this work is to derive the rotation dependence of stellar-wind characteristics and compare them with observations.

The remainder of this paper is organized as follows.
In Section 2, we summarize the overview of the the model in this work, including assumptions, basic equations, parameters, and numerical schemes.
The numerical results are discussed in Section 3.
The energetics of the stellar wind is discussed in Section 4, focusing on the mass-loss saturation and wave energetics.
We discuss the overall results of our work in Section 5.

\section{Model}

\subsection{Overview}

\begin{figure*}[t]
\centering
\includegraphics[width=160mm]{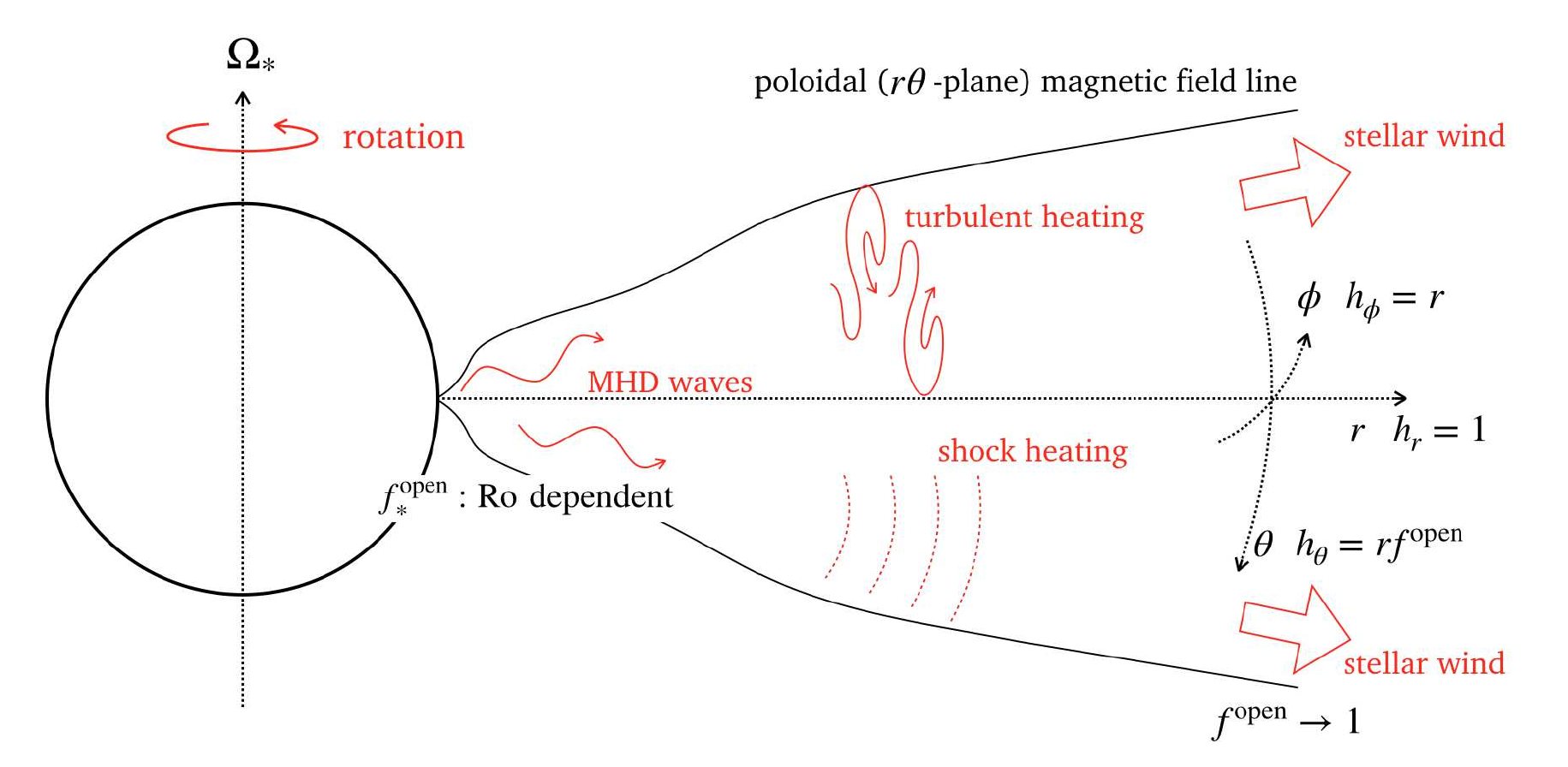}
 \caption{
 	A schematic picture of the stellar wind geometry used in this study.
	Shown by black symbols and characters are the numerical settings.
	Red symbols and characters refer to the physical processes considered in this work.
 }     
\label{fig:coordinate_physics}
\vspace{1em}
\end{figure*}

We simulate equatorial stellar winds that extend from the stellar photosphere to beyond the fast magnetosonic point.
Our model is based on the magnetohydrodynamic equations including gravity, thermal conduction, and radiative cooling.
For simplicity, and to reduce the numerical cost, we assume a one-dimensional geometry and axisymmetry.
We therefore include the turbulent dissipation of Alfv\'en waves, which is a multi-dimensional effect, phenomenologically.

Several theoretical models explain the solar wind based on Alfv\'en-wave heating and acceleration \citep{Suzuk06a,Cranm07,Holst14,Shoda19,Revil20}.
We extend this Alfv\'en-wave modeling to winds from low-mass stars.
In addition to Alfv\'en waves, for fast-rotators the magneto-centrifugal force can further accelerate the wind \citep{Belch76,Sakur85,Revil16,Johns17}.
In this work, we account for both effects and conventionally call our models ``Alfv\'en-wave driven magnetic rotator winds''.
For simplicity, we fix the mass, luminosity, and metallicity of the star with solar values ($M_\ast = M_\odot, \ L_\ast = L_\odot, \ Z_\ast = Z_\odot$) and focus our interest on the rotation dependence.

A key factor in our model (and in the theory of magnetic braking in general) is the filling factor of open magnetic flux, defined as
\begin{align}
    f^{\rm open}(r) = \frac{\Phi_{\rm open}}{4 \pi r^2 \left| B_r (r) \right|},
\end{align}
where $\left| B_r (r) \right|$ is the unsigned radial magnetic field (averaged over solid angle) and $\Phi_{\rm open}$ is the unsigned open magnetic flux.
Note that $\Phi_{\rm open}$ is constant in $r$ and is fixed for each simulation.
Magnetic field lines eventually become open, due to their advection in the stellar wind, therefore $f^{\rm open} (r \to \infty) = 1$.
At the stellar surface, $f^{\rm open}$ is generally much lower than unity.
For the solar surface, $f^{\rm open}$ is typically $10^{-3}$ \citep{Cranm17a}.
The radial increase of $f^{\rm open}$ as a function of radius, i.e. the (super-radial) expansion of open magnetic field line, needs to be accounted for.

To follow the (super-radial) magnetic field expansion in our one-dimensional geometry, we make use of a field-aligned coordinate system \citep{Hollw82a,Kudoh99,Suzuk05}.
To be consistent with axisymmetry, all the super-radial expansion is attributed to the poloidal ($r$ and $\theta$) components.
The scale factors (that reflect the degree of expansion of magnetic flux in each direction) of the corresponding curvilinear coordinate system are given as
\begin{align}
  h_r = 1, \ \ \ h_\theta = rf^{\rm open}, \ \ \ h_\phi = r. \label{eq:scale_factors} 
\end{align}
For simplicity, $\theta$ and $\phi$ components are attributed to (Alfv\'en) waves and rotation, respectively.
By this simplification, the polarization of Alfv\'en waves is restricted to be linear.
However, this restriction is unlikely to affect the conclusion because the wind structure and dynamics are weakly affected by the imposed polarization \citep{Suzuk06a}.

Although stellar wind outflows are far from symmetric \citep[e.g.][]{Holst14},
once the flux-tube expansion is appropriately prescribed by $f^{\rm open} (r)$, 
our one-dimensional model is expected to give a reasonable estimation of wind parameters.
In the wind acceleration region and below, 
due to the low-beta nature of the corona, 
the flux-tube expansion is essentially set by the global magnetic field and is not affected by the wind dynamics.
Because the interactions between flux tubes are likely insignificant, in wave-driven winds,
each flux tube behaves independently.
Indeed, the three-dimensional structure of the solar wind is well reproduced from an ensemble of one-dimensional flux-tube models \citep{Pinto17}.
Thus, by implementing a representative flux-tube expansion, 
we can reliably recover averaged properties of the stellar wind.

An overview of our model geometry is detailed in Figure \ref{fig:coordinate_physics}.
An equatorial magnetic flux tube is located on the stellar surface and expands super-radially into the interplanetary space.
MHD waves propagate along the background flux tube and partially dissipate in the atmosphere.

\subsection{Basic equations}
We assume that the system is one-dimensional ($\partial/\partial \theta = \partial / \partial \phi = 0$) 
and the scale factors are given by Eq. (\ref{eq:scale_factors}). 
The MHD equations are then written as follows (see Appendix \ref{app:derivation} for derivation):
\begin{align}
  &\frac{\partial}{\partial t} \rho + \frac{1}{r^2 f^{\rm open}} \left( \rho v_r r^2 f^{\rm open}\right) = 0, \label{eq:basic_ro} \\
  &\frac{\partial}{\partial t} \left( \rho v_r \right) + \frac{1}{r^2 f^{\rm open}} \frac{\partial}{\partial r}\left[ \left( \rho v_r^2 + p_T \right) r^2 f^{\rm open} \right] \nonumber \\
  & \hspace{1em} = - \rho \frac{G M_\ast}{r^2} + p \frac{d}{dr} \ln \left( r^2 f^{\rm open} \right) \nonumber \\
  & \hspace{1em} + \rho v_\theta^2 \frac{d}{dr} \ln \left( r f^{\rm open} \right) + \frac{1}{r} \rho v_\phi^2 - \left( \frac{B_\theta^2}{8 \pi} - \frac{B_\phi^2}{8 \pi} \right) \frac{d}{dr} \ln f^{\rm open}, \\ \label{eq:basic_vr}
  &\frac{\partial}{\partial t} \left( \rho v_\theta \right) + \frac{1}{r^2 f^{\rm open}} \frac{\partial}{\partial r} \left[ \left( \rho v_r v_\theta - \frac{1}{4 \pi} B_r B_\theta \right) r^2 f^{\rm open} \right] \nonumber \\
  & \hspace{1em} = \left( \frac{B_r B_\theta}{4 \pi} - \rho v_r v_\theta \right) \frac{d}{dr} \ln \left(r f^{\rm open} \right) + \rho D^{\rm turb}_v, \\
  &\frac{\partial}{\partial t} \left( \rho v_\phi \right) + \frac{1}{r^2 f^{\rm open}} \frac{\partial}{\partial r} \left[ \left( \rho v_r v_\phi - \frac{1}{4 \pi} B_r B_\phi \right) r^2 f^{\rm open} \right] \nonumber \\
  & \hspace{1em} = \left( \frac{B_r B_\phi}{4 \pi} - \rho v_r v_\phi \right)/r,  \\
  &\frac{1}{r^2 f^{\rm open}} \frac{d}{dr} \left( B_r r^2 f^{\rm open} \right) = 0, \\
  &\frac{\partial}{\partial t} B_\theta + \frac{1}{r^2 f^{\rm open}} \frac{\partial}{\partial r} \left[ \left( v_r B_\theta - v_\theta B_r \right) r^2 f^{\rm open}  \right] \nonumber \\
  & \hspace{1em} = \left( v_r B_\theta - v_\theta B_r \right) \frac{d}{dr} \ln \left( r f^{\rm open} \right) + \sqrt{4 \pi \rho} D^{\rm turb}_b, \\
  &\frac{\partial}{\partial t} B_\phi + \frac{1}{r^2 f^{\rm open}} \frac{\partial}{\partial r} \left[ \left( v_r B_\phi - v_\phi B_r \right)  r^2 f^{\rm open} \right] \nonumber \\
  & \hspace{1em} = \left( v_r B_\phi - v_\phi B_r \right) /r, \\
  &\frac{\partial}{\partial t} e + \frac{1}{r^2 f^{\rm open}} \frac{\partial}{\partial r} 
  \left[ \left( \left( e + p_T \right) v_r - \frac{B_r}{4 \pi} \left( \vec{v}_\perp \cdot \vec{B}_\perp \right) + F_C \right)  r^2 f^{\rm open} \right] \nonumber \\
  & \hspace{1em} = - \rho v_r  \frac{G M_\ast}{r^2} - Q_R, \label{eq:basic_en}
\end{align} 
where
\begin{align}
	e = \frac{p}{\gamma-1} + \frac{1}{2} \rho v^2 + \frac{B_\perp^2}{8 \pi},  \ \ \ p_T = p + \frac{B_\perp^2}{8 \pi}.
\end{align}
These are closed by  the equation of state.
\begin{align}
	p = c_1 \rho T, \ \ \ c_1 = 1.36 \times 10^8 {\rm \ erg \ g^{-1} \ K^{-1}}
\end{align}
where the value of $c_1$ is consistent with the equation of state of a fully-ionized plasma that contains a few percent helium (alpha particle) by number.
$D_v^{\rm turb}$ and $D_b^{\rm turb}$ represent the rate of turbulent dissipation of Alfv\'en wave per unit momentum \citep{Shoda18a}, which will be discussed in Section \ref{sec:alfven_wave_turbulence}.
$F_C$ and $Q_R$ are the conductive flux and radiative cooling rate, respectively.
 
We employ a Spitzer-H\"arm type of thermal conductive flux \citep{Spitz53}, with a quenching term that works at radial distances typically greater than $5R_\odot$:
\begin{align}
  F_C = - \min \left( 1, \frac{\rho}{\rho_C} \right) \frac{B_r}{\left| \vec{B} \right|} \kappa_0 T^{5/2} \frac{dT}{dr},
\end{align}
where we set $\kappa_0 = 10^{-6} \ {\rm erg \ cm^{-1} \ s^{-1} \ K^{-7/2}}$ and $\rho_C = 10^{-20} \ {\rm g \ cm^{-3}}$.
The quenching term comes from the saturation of heat flux in the interplanetary space \citep{Salem03,Bale013}.
Although this quenching is known to be overestimated, 
it is unlikely to affect the numerical result as it generally occurs beyond the sonic point.

The radiative cooling is a combination of different types:
\begin{align}
  Q_R = Q_{R,{\rm thck}} \xi_1 + Q_{R,{\rm thin}} \left( 1 - \xi_1 \right),
\end{align}
where $Q_{R,{\rm thck}}$ and $Q_{R,{\rm thin}}$ stand for the optically thick and thin radiative losses, respectively.
The switching parameter $\xi_1$ mimics the optical depth, which takes $\xi_1 \approx 1$ in the photosphere and $\xi_1 \approx 0$ in the corona.
Here we assume the following expression for $\xi_1$:
\begin{align}
  \xi_1 = \max \left(0, 1- \frac{p_{\rm chr}}{p} \right), 
\end{align}
where $p_{\rm chr} = 1 {\rm \ dyn \ cm^{-2}}$.

In the photosphere, where the optical depth is large, the balance between radiative heating and cooling keeps the temperature almost fixed.
For this reason, following \citet{Gudik05}, we approximate the optically thick cooling by an exponential cooling:
\begin{align}
  Q_{R,{\rm thck}} &= \frac{1}{\tau_{\rm thck}} \left( e_{\rm int} - e_{\rm int, ref} \right), \\
  \tau_{\rm thck} &= 0.1  \left( \frac{\rho}{\overline{\rho}_\ast} \right)^{-5/3} {\rm s},
\end{align}
where $\overline{\rho} = 10^{-7} {\rm \ g \ cm^{-3}}$ is the mean surface density, and $e_{\rm int,ref}$ is the internal energy at a given reference temperature that mimics the radiation-balanced profile \citep[e.g. Figure 4 in][]{Cranm19}.
$Q_{R,{\rm thck}}$ only works near the surface because of the rapid increase of $\tau_{\rm thck}$ with height.

Following \citet{Iijim16}, the optically-thin cooling function is composed of two different contributions.
In the chromosphere, we employ the radiative cooling function given by \citet{Goodm12} ($Q_{\rm GJ}$), while in the corona, the optically thin cooling function taken from \citet{Rempe17} is used.
These two functions are smoothly connected as a function of temperature using:
\begin{align}
  Q_{R,{\rm thin}} &= Q_{\rm GJ} (\rho,T) \xi_2 +  n_p n_e \Lambda(T) (1-\xi_2), \\
  \xi_2 &=  \max \left(0,  \min \left(1, \frac{T_{\rm TR} - T}{\Delta T } \right) \right), 
\end{align}
where $T_{\rm TR} = 15000 {\rm \ K}$ and $\Delta T = 5000 {\rm \ K}$. 

\begin{table*}[t!]
\centering
  \begin{tabular}{c c c c c c c} 
    \begin{tabular}{c} $P_{\rm rot}$ \hspace{1em} \\ ${\rm \ [day]}$ \hspace{1em} \end{tabular} & \begin{tabular}{c} $f^{\rm open}_\ast$ \hspace{1em} \\ ${[10^{-3}]}$ \hspace{1em} \end{tabular} & \begin{tabular}{c} $\dot{M}_w$ \hspace{1em} \\ $[10^{-14}  \ M_\odot \ {\rm yr^{-1}}]$ \hspace{1em} \end{tabular} & \begin{tabular}{c} $r_{\rm A}$ \hspace{1em} \\ $[R_\odot]$ \hspace{1em} \end{tabular} & \begin{tabular}{c} $v_{r,A}$ \hspace{1em} \\ ${\rm [10^2 \ km \ s^{-1}}]$ \hspace{1em} \end{tabular} & \begin{tabular}{c} $\tau_w$ \hspace{1em} \\ $[10^{30} \ {\rm erg}]$ \hspace{1em} \end{tabular} \rule[-4.5pt]{0pt}{16pt}  \\ \hline \hline
    48 &  0.466  &  1.01   &  11.4  &  2.14  &  0.61 \rule[-4.5pt]{0pt}{16pt} \\
    40 &  0.580  &  1.26   &  12.2  &  2.33  &  1.03 \rule[-4.5pt]{0pt}{16pt} \\ 
    32 &  0.758  &  1.58   &  13.3  &  2.62  &  1.96 \rule[-4.5pt]{0pt}{16pt} \\ 
    24 &  1.07   &  2.00   &  15.6  &  3.08  &  4.48 \rule[-4.5pt]{0pt}{16pt} \\
    20 &  1.33   &  2.23   &  17.5  &  3.40  &  7.54 \rule[-4.5pt]{0pt}{16pt} \\ 
    16 &  1.74   &  2.49   &  20.5  &  3.80  &  14.4 \rule[-4.5pt]{0pt}{16pt} \\ 
    12 &  2.46   &  2.79   &  25.3  &  4.41  &  33.0 \rule[-4.5pt]{0pt}{16pt} \\
    10 &  3.06   &  2.87   &  29.6  &  4.81  &  56.0 \rule[-4.5pt]{0pt}{16pt} \\ 
    8  &  4.00   &  3.05   &  35.8  &  5.34  &  108  \rule[-4.5pt]{0pt}{16pt} \\ 
    6  &  5.65   &  3.25   &  46.2  &  5.95  &  258  \rule[-4.5pt]{0pt}{16pt} \\
    5  &  7.03   &  3.41   &  54.2  &  6.57  &  433  \rule[-4.5pt]{0pt}{16pt} \\ 
    4  &  9.19   &  3.36   &  64.9  &  7.37  &  825  \rule[-4.5pt]{0pt}{16pt} \\ 
    3  &  13.0   &  3.28   &  80.2  &  9.81  &  1650 \rule[-4.5pt]{0pt}{16pt} \\ 
    2  &  21.1   &  3.09   &  107   &  16.5  &  3890 \rule[-4.5pt]{0pt}{16pt} \\ \hline
  \end{tabular}
  \vspace{1.0em}
  \caption{Summary of the input and output parameters of our simulations.
  The first two columns correspond to the input parameters (rotation period and open-flux filling factor), while the last four columns show the output parameters (mass-loss rate, Alfv\'en radius, Alfv\'en-point wind velocity, angular momentum loss rate).}
  \vspace{0.5em}
  \label{table:parameters}
\end{table*}

\subsection{Open-flux filling factor}

There is a strong relationship between the open magnetic flux, $\Phi_{\rm open}$, and the strength of magnetic braking \citep{Vidot12,Vidot14b,Revil15}.
The magnetic flux conservation yields
\begin{align}
  \Phi_{\rm open} = 4 \pi R_\ast^2 B_\ast f^{\rm open}_\ast \approx 4 \pi R_\ast^2 B_{{\rm eq},\ast}  f^{\rm open}_\ast,
\end{align}
where $B_\ast$ is the characteristic field strength at the photosphere, and can be approximated by the equipartition value $B_{{\rm eq},\ast}$ that represents equal gas and magnetic pressures \citep{Cranm11}.
We assume that the stellar surface is divided into two areas: one with zero field and the other with equipartition field.
Indeed, the Sun's photospheric magnetic field is observed to be spatially localized, exhibiting a nearly equipartition value \citep{Tsune08}.
Under this assumption, $f^{\rm open}_\ast$ represents the fraction of the stellar surface covered by open magnetic flux \citep{Saar001,Reine09}.
We note that the "open-flux filling factor" does not stand for the fraction of the open flux to the total flux $f^{\rm open}_\ast/(f^{\rm open}_\ast + f^{\rm closed}_\ast)$, 
where $f^{\rm closed}_\ast$ is the fraction of the stellar surface covered by closed magnetic flux.
We also note that, for the solar wind, $f^{\rm open}_\ast$ (or,equivalently, the expansion factor) plays a role in determining the wind speed \citep{Wang090,Arge000,Fujik15}.

Unfortunately, there is no established way to determine $f^{\rm open}_\ast$ from the photospheric magnetic field, even for the Sun.
For example, the widely used Potential Field Source Surface model \citep{Schat69} consistently underestimates the open magnetic flux observed by in-situ spacecraft, which is referred to as the open-flux problem \citep{Linke17}. 
However, it is thought that the dipolar magnetic field is the most significant contributor to the open magnetic flux \citep{Revil15,See0018}.
Recently, \citet{See0019b} showed that, for most stars (especially with low Rossby numbers),
the dipolar magnetic field is sufficient to determine the angular momentum loss rate. 
For these reasons, we simply assume that the open magnetic flux is proportional to the surface-averaged unsigned dipolar magnetic field $\langle B_{\rm dip} \rangle$:
\begin{align}
  f^{\rm open}_\ast \propto \frac{\langle B_{\rm dip} \rangle}{B_{{\rm eq},\ast}}.
\end{align}
In this work, we assume a power-law relation for $f^{\rm open}_\ast$ as
\begin{align}
  f^{\rm open}_\ast = f^{\rm open}_\odot \left( \frac{{\rm Ro}}{{\rm Ro}_\odot} \right)^{-1.2}
  \approx 10^{-3} \left( \frac{P_{\rm rot}}{P_{{\rm rot},\odot}} \right)^{-1.2}, \label{eq:fopen_thiswork}
\end{align}
where we use $f^{\rm open}_\odot \approx 10^{-3}$ as the the solar value of $f^{\rm open}_\ast$.
Note that we only consider main-sequence Sun-like stars ($M_\ast = M_\odot$, $R_\ast = R_\odot$, $Z_\ast = Z_\odot$)
and therefore the Rossby number is a function of rotation rate only.
Our implementation is in line with the derivation presented in \citet{See0019b}, who showed 
\begin{align}
  \langle B_{\rm dip} \rangle \propto {\rm Ro}^{-1.3 \pm 0.1}, \label{eq:dipolar_see2019}
\end{align}
based on a statistical analysis of Zeeman-Doppler imaging (ZDI) observations.
The actual dependence of $\langle B_{\rm dip} \rangle$ on ${\rm Ro}$ could be weaker because the ZDI observation tends to underestimate the magnetic field strength, especially when the field is weak \citep{See0020}.
This supports our assumption that $\langle B_{\rm dip} \rangle$ depends more weakly on ${\rm Ro}$ than observation by \citet{See0019b}.
In future, a rotation-dependent correction factor to the ZDI-based $\langle B_{\rm dip} \rangle$ values could be used.
We summarize the input and output parameters of our simulations in Table \ref{table:parameters}.

We assume a two-step super-radial expansion of the magnetic field line: one expansion occurs in the chromosphere and the other in the corona \citep{Cranm05}.
To implement such two-step expansion, we need to set the filling factor between the two expansion regions (at the coronal base),
which we denote $f^{\rm open}_{\rm cor}$.
Following \citet{Cranm11}, we simply assume a power-law relation between $f^{\rm open}_\ast$ and $f^{\rm open}_{\rm cor}$ as
\begin{align}
  f^{\rm open}_{\rm cor} = \left( f^{\rm open}_\ast \right)^{\theta_B}, 
\end{align}
where we use $\theta_B = 1/3$ as a reference value.

Once $f^{\rm open}_\ast$ and $f^{\rm open}_{\rm cor}$ are given, we set the radial profile of $f_{\rm open} (r)$ as
\begin{subequations}
\begin{align}
	&f^{\rm open} (r) = f^{\rm open}_\ast f^{\rm exp}_1 (r) f^{\rm exp}_2 (r),  \\
	&f^{\rm exp}_1 (r) = \min \left[ f^{\rm open}_{\rm cor}/f^{\rm open}_\ast, \exp \left( \frac{r-R_\ast}{2 h_{\rm exp}} \right) \right], \label{eq:fopen1} \\
	&f^{\rm exp}_2 (r) = \frac{\mathcal{F}(r) + f^{\rm open}_{\rm cor} + \mathcal{F}(R_\ast) \left( f^{\rm open}_{\rm cor} - 1 \right)}{f^{\rm open}_{\rm cor} \left( \mathcal{F}(r)+1 \right)},
  \label{eq:fopen2} 
\end{align}
\end{subequations}
where $\mathcal{F}(r) = \exp \left( \frac{r - r_{\rm exp} }{\sigma_{\rm exp}} \right)$.
$f^{\rm exp}_1$ and $f^{\rm exp}_2$ represent the degree of flux-tube expansion in the chromosphere and corona, respectively.
We assume that, in the stellar chromosphere, the flux tube expands so that the plasma beta is fixed until $f^{\exp}_1 = f^{\rm open}_{\rm cor}/f^{\rm open}_\ast$ \citep{Tsune08}.
Such an expansion is approximately realized by setting the scale height $h_{\rm exp}$ as
\begin{align}
	h_{\rm exp} = a_\ast^2 / g_\ast,
\end{align}
where $a_\ast$ and $g_\ast$ are the sound speed and gravitational acceleration at the stellar surface, respectively.
Here we assume that the pressure scale height of the chromosphere is similar to the photospheric value.
For the coronal expansion, we follow the formulation of \citet{Kopp076}, with $r_{\rm exp} / R_\ast = 1.2$ and $\sigma_{\rm exp} / R_\ast = 0.3$.

\subsection{Alfv\'en wave turbulence}\label{sec:alfven_wave_turbulence}
Broadband energy spectra observed in the solar wind indicate that the solar wind is at least partially heated by turbulence \citep{Colem68,Belch71a,Podes07,Chen020}.
In fact, in the outer heliosphere, the observed turbulent dissipation accounts for the required heating rate of the solar wind \citep{Carbo09}.
Although it is still unclear how the solar wind is energized in and below the acceleration region,
it is straightforward to assume that the heating process should be similar to what we observe in the distant solar wind;
i.e. plasma is heated by turbulence in the solar atmosphere. 
Alfv\'en wave turbulence is a promising candidate of such a heating mechanism. 
It is a type of MHD turbulence that is driven by the collision of bi-directional Alfv\'en waves or Els\"asser variables \citep{Kraic65,Dobro80,Howes13}. It is likely to develop in the stellar atmosphere (corona) and wind because the reflection of Alfv\'en waves therein naturally gives rise to wave-wave collisions \citep{Matth99,Dmitr02}.
Alfv\'en-wave turbulence is now regarded as one of the most dominant heating processes in coronal holes and the fast solar wind \citep{Verdi07,Cranm07,Perez13,Balle16,Shoda19},
in coronal loops \citep{Balle11,Verdi12b}, and in the chromosphere \citep{Verdi07,Balle11}.
Note, however, that other processes such as mode conversion \citep{Moriy04,Suzuk05,Antol08},
parametric decay instability \citep{Suzuk06a,Tener13,DelZa15,Shoda18d,Revil18}, and phase mixing \citep{Heyva83,Magya17} 
are also likely to be important.

Without any additional terms, one-dimensional models cannot deal with Alfv\'en wave turbulence,
because it is a multi-dimensional process.
To model the Alfv\'en wave turbulence without expensive numerical cost, phenomenological treatments have been proposed \citep{Hossa95,Dmitr02}.
These models have been validated in previous solar wind simulations \citep[e.g.][]{Balle16}.
Following \citet{Shoda18a}, we introduce a phenomenological model of turbulent dissipation as
\begin{subequations}
\begin{align}
  D^{\rm turb}_v &= \frac{c_d}{4 \lambda_\perp} \left( \left| z_\theta^+ \right| z_\theta^- + \left| z_\theta^- \right| z_\theta^+ \right),  \label{eq:Dturb_v} \\
  D^{\rm turb}_b &= \frac{c_d}{4 \lambda_\perp} \left( \left| z_\theta^+ \right| z_\theta^- - \left| z_\theta^- \right| z_\theta^+ \right), \label{eq:Dturb_b} 
\end{align}
\end{subequations}
where $\lambda_\perp$ is the perpendicular correlation length and $z_\theta^\pm$ are Els\"asser variables \citep{Elsas50}:
\begin{align}
  z_\theta^\pm = v_\theta \mp B_\theta / \sqrt{4 \pi \rho}. \label{eq:elsasser}
\end{align}
We assume that the correlation length increases with the flux-tube radius:
\begin{align}
  \lambda_\perp = \lambda_{\perp,\ast} \sqrt{\frac{B_\ast}{B_r}}.
\end{align}
In the photosphere, Alfv\'enic fluctuations are localized in the inter-granular lanes where magnetic flux is concentrated \citep{Balle98,Balle11,Chitt12}.
For this reason, we set the photospheric correlation length of Alfv\'en-wave turbulence as the typical width of inter-granular lane:
\begin{align}
  \lambda_{\perp,\ast} = 100 {\rm \ km}.
\end{align}
For the value of $c_d$ in Eq.s (\ref{eq:Dturb_v}) and (\ref{eq:Dturb_b}), following \citet{Shoda18a}, we set
\begin{align}
  c_d = 0.1,
\end{align}
which is supported by both a reduced-MHD simulation \citep{Balle17} and a shell-model calculation \citep{Verdi19b}.
However, the best choice of $c_d$ remains controversial as one reduced-MHD calculation by \citet{Chand19} shows $c_d \sim 1$.
The uncertainty in $c_d$ is not a key issue in this work because the stellar wind parameters appear to weakly depend on the value of $c_d$ \citep[see][]{Shoda18a}.

\subsection{Simulation domain and boundary condition}
We solve the basic equations from the photosphere ($r=R_\ast$) to the outer boundary of the stellar wind ($r=r_{\rm out}$).
The extent of the simulation domain changes depending on $P_{\rm rot}$, such that the $r_{\rm out}$ is always beyond the fast-magnetosonic point.
For example, we set $r_{\rm out} / R_\ast = 100$ when $P_{\rm rot} = 24 {\rm \ day}$ and $r_{\rm out} / R_\ast = 690$ when $P_{\rm rot} = 2 {\rm \ day}$.

The spatial resolution of the simulation domain is inhomogeneous.
Below $r = 1.02 R_\ast$ the grid size $\Delta r$ is fixed to $\Delta r = 20 {\rm \ km}$ independent of $\Omega_\ast$.
$\Delta r$ increases with $r$ as a power-law of $r$ above $r = 1.02 R_\ast$ until it reaches the maximum value, $\Delta r_{\rm max}$.
To resolve Alfv\'en waves without large numerical cost, we increase $\Delta r_{\rm max}$ with rotation rate, $\Omega_\ast$,
because stellar wind speed and Alfv\'en velocity are larger in faster rotators.
Specifically, $\Delta r_{\rm max} = 4 \times 10^3 {\rm \ km}$ for $P_{\rm rot} = 48 {\rm \ day}$
and $\Delta r_{\rm max} = 10^4 {\rm \ km}$ for $P_{\rm rot} = 2 {\rm \ day}$.

Beyond the outer boundary, $r_{\rm out}$, a marginal simulation domain is set with gradually increasing grid size.
Any numerical errors in the marginal region are unlikely to affect the simulation result since the outer boundary is always beyond the fast magnetosonic point, where physical fluctuations cannot propagate back into the simulation domain.

Values evaluated at the inner boundary are denoted with the subscript $\ast$, and are given as follows.
Fixed boundary conditions are imposed for $T$, $v_\phi$, $B_r$, and $B_\phi$:
\begin{align}
	\begin{split}
		T_\ast &= \frac{p_\ast}{c_1 \rho_\ast} = 6000 {\rm \ K}, \ \ \ v_{\phi,\ast} = R_\ast \Omega_\ast \\ 
		B_{r,\ast} &= B_{{\rm eq},\ast} = 1300 {\rm \ G}, \ \ \ B_{\phi,\ast} = 0, 
	\end{split}
\end{align}
Note that $B_{{\rm eq},\ast}$ is assumed to be constant with respect to $\Omega_\ast$.
Because the photospheric motion exhibits much smaller timescale than the rotation,
the local property of the photosphere is independent from rotation rate.

To inject MHD waves at the photosphere, we impose time-dependent boundary conditions for density, velocity and perpendicular magnetic field.
Fluctuations of density and radial velocity are given as
\begin{align}
	\rho_\ast = \overline{\rho}_\ast \left( 1 + \frac{v_{r,\ast}}{a_\ast} \right)
\end{align}
where
\begin{align}
	\overline{\rho}_\ast = 10^{-7} {\rm \ g \ cm^{-3}}, \ \ a_\ast = \sqrt{c_1  T_\ast}. 
\end{align}
The time dependent radial (vertical) velocity $v_{r,\ast}$ has a broadband spectrum of
\begin{align}
	v_{r,\ast} \propto \sum_{N=0}^{10} \sin \left( 2 \pi f^l_N t + \phi^l_N \right) / \sqrt{2 \pi f^l_N},
\end{align}
where the $\phi^l_N$ is a random phase and the wave frequency $f^l_N$ ranges in $3.33 \times 10^{-3} {\rm \ Hz} \le f^l_N \le 3.33 \times 10^{-2} {\rm \ Hz}$.
The lower limit of $f^l_N$ is set to be the cut-off frequency of acoustic waves at the stellar surface.
The amplitude of $v_{r,\ast}$ is set so that the root-mean-squared amplitude of upward acoustic waves is $0.9 {\rm \ km \ s^{-1}}$.
Considering the downward wave contribution, the root-mean-squared vertical velocity at the surface is approximately $1.3 {\rm \ km \ s^{-1}}$, consistent with solar observations \citep{Oba0017,Ishik20}.

The $\theta$-component of the velocity and magnetic field are given in terms of Els\"asser variables from Eq.(\ref{eq:elsasser}).
We impose a zero-derivative boundary condition on $z_{\theta}^-$ such that reflected Alfv\'en waves can be absorbed through the bottom boundary:
\begin{align}
	\left. \frac{\partial}{\partial r} z_{\theta}^- \right|_\ast = 0.
\end{align}
Like $v_{r,\ast}$, upward Els\"asser variable $z^+_{\theta,\ast}$ is given with a broadband spectrum as follows.
\begin{align}
	z^+_{\theta,\ast} \propto \sum_{N=0}^{20} \sin \left( 2 \pi f^t_N t + \phi^t_N \right) / \sqrt{2 \pi f^t_N},
\end{align}
where $\phi^t_N$ is a random phase and $f^t_N$ ranges in $1.00 \times 10^{-3} {\rm \ Hz} \le f^t_N \le 1.00 \times 10^{-2} {\rm \ Hz}$.
The lower and upper limits of this frequency range approximates the turn-over timescale in granules and intergranular lanes \citep{Hirzb99}.
The amplitude is tuned so that the root-mean-squared amplitude of upward Alfv\'en waves is $1.2 {\rm \ km \ s^{-1}}$, which yields the root-mean-squared photospheric transverse velocity of $1.7 {\rm \ km \ s^{-1}}$.
This value is consistent with observations of the solar surface convection \citep{deWij08,Oba0020}.
There is evidence that the imposed spectrum of transverse waves may affect the dynamics of the resulting stellar wind \citep{Shoda18d}.
Therefore, in the future, simulations should be performed with a self-consistent convection zone  \citep[e.g.][]{Rempe17} to remove the uncertainty in the wave generation process.

\begin{figure*}[t]
\centering
\includegraphics[width=180mm]{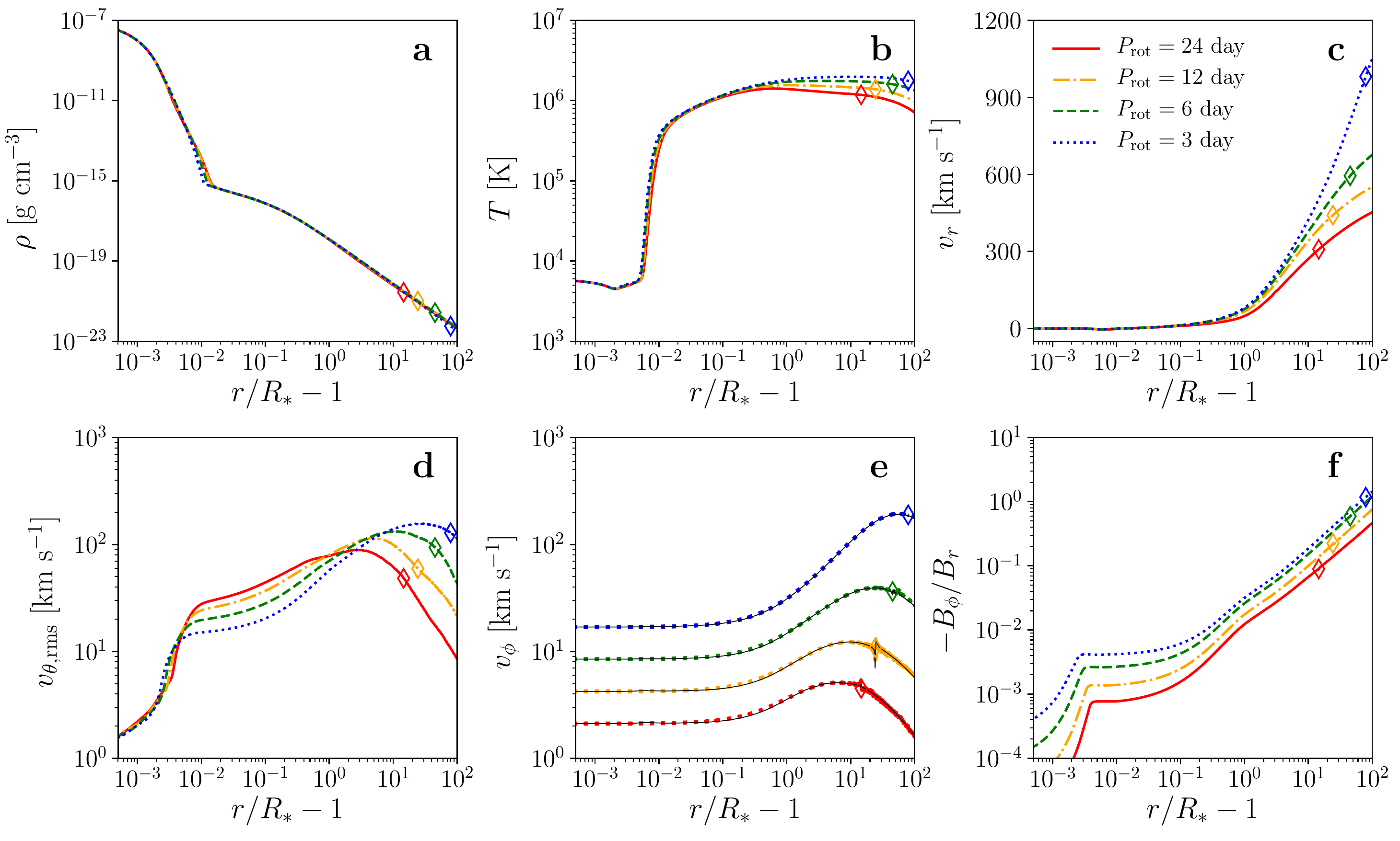}
\caption
{
  Comparison of the quasi-steady state solutions after time averaging.
  The four lines correspond to
  $P_{\rm rot} = 24 {\rm \ days}$ (red solid line),
  $P_{\rm rot} = 12 {\rm \ days}$ (orange dash-dotted line),
  $P_{\rm rot} = 6 {\rm \ days}$ (green dashed line) and
  $P_{\rm rot} = 3 {\rm \ days}$ (blue dotted line) respectively.
  {\bf a}: mass density $\rho$,
  {\bf b}: temperature $T$,
  {\bf c}: radial velocity $v_r$,
  {\bf d}: root-mean-squared wave amplitude $v_{\theta,{\rm rms}}$,
  {\bf e}: rotation velocity $v_\phi$,
  {\bf f}: field inclination $- B_\phi / B_r$.
  Thin black lines in Panel {\bf e} indicate the Weber-Davis solution and, for better visualization, the wind simulations are shown with dotted lines (only in Panel {\bf e}).
  Diamonds indicate the Alfv\'en point.
}     
\label{fig:comparison_time_averaged}
\vspace{1em}
\end{figure*}

\section{Trends in the wind simulations}

\subsection{Overview of rotation dependence}

In Figure \ref{fig:comparison_time_averaged}, we show the time-averaged radial profiles of the simulated winds with various rotation rates: $P_{\rm rot} = 24 {\rm \ day}$ (red solid line), $12 {\rm \ day}$ (orange dash-dotted line), $6 {\rm \ day}$ (green dashed line) and $3 {\rm \ day}$ (blue dotted line).
To eliminate the influence of initial conditions, time averaging is conducted after the system reaches a quasi-steady state that is independent from the choice of initial condition.
The averaging time is typically $1.2 \times 10^5 {\rm \ s} \approx 1.39 {\rm \ day}$.
%The panels in Figure \ref{fig:comparison_time_averaged} represents the
%mass density ($\rho$, Panel {\bf a}),
%temperature ($T$, Panel {\bf b}),
%radial velocity ($v_r$, Panel {\bf c}),
%wave amplitude ($v_\theta$, Panel {\bf d}),
%rotation velocity ($v_\phi$, Panel {\bf e}) and normalized azimuthal magnetic filed ($-B_\phi / B_r$, Panel {\bf f}) respectively.
Note, we show the root-mean-squared value for the wave amplitude $v_\theta$.

Panel {\bf e} directly reflects the different rotation velocities used in our simulations; $v_\phi$ increases with rotation rate.
Up to a certain height, the wind co-rotates with the stellar surface.
The co-rotation breaks up below the Alfv\'en radius, and $v_\phi$ in turn begins to decrease.
This behavior is consistent with the Weber-Davis solution, $v_{\phi,{\rm WD}}$, which predicts
\begin{align}
	\begin{split}
 		 v_{\phi,{\rm WD}} &\approx r \Omega_\ast \ \ \ \ \ \ \ (r \ll r_A ), \\
 		 v_{\phi,{\rm WD}} &\approx r_A^2 \Omega_\ast / r \ \ (r \gg r_A ).
	 \end{split}
\end{align}
In fact, the radial profile of $v_\phi$ almost perfectly coincides with the Weber-Davis solution (thin black lines).
Similarly, Panel {\bf f} shows that $-B_\phi / B_r$ increases with $r$, typically $-B_\phi / B_r \propto r$ far away from the star.
This is also consistent with Weber-Davis solution that predicts the Parker-spiral relation \citep{Parke58}.
\begin{align}
  - B_{\phi,{\rm WD}} / B_r \approx r \Omega_\ast / v_{r,\infty} \ \ (r \gg r_A ). 
\end{align}

Diamonds on each line indicate the Alfv\'en point.
The Alfv\'en point occurs at larger radii as the rotation rate increases.
This is a natural consequence of larger open magnetic flux and the larger coronal Alfv\'en speed of the faster rotators.
As shown in Table 1, the largest Alfv\'en radius exceeds the mean orbital radius of Mercury ($\sim 83 R_\odot$).
This indicates that Mercury was possibly subject to magnetic star-planet interactions with the young Sun \citep[see e.g.,][]{Strug14,Folso20}.

\begin{figure}[t]
\centering
\includegraphics[width=72mm]{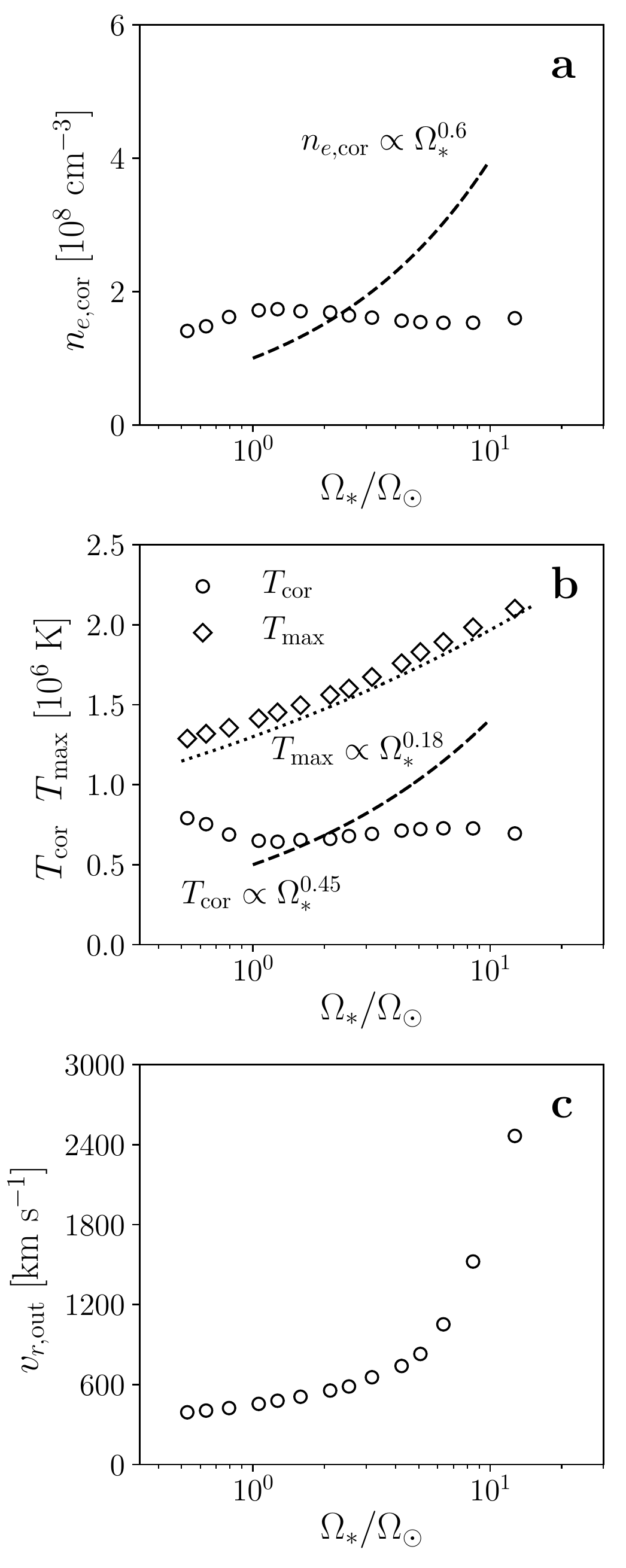}
\caption
{
  $\Omega_\ast$-dependences of stellar wind parameters.
  {\bf a}: coronal-base electron number density, $n_{e,{\rm cor}}$. 
  Also shown by dashed line is the empirical scaling of $n_{e,{\rm cor}} \propto \Omega_\ast^{0.6}$ \citep{Ivano03}.
  {\bf b}: coronal-base temperature ($T_{\rm cor}$, circles) and maximum temperature ($T_{\rm max}$, diamonds).
  Dashed line is the observational single-power-law relation \citep{Fionn18}.
  Dotted line is the power-law fitting between $T_{\rm max}$ and $\Omega_\ast$.
  {\bf c}: termination velocity of stellar wind.
 }     
 \label{fig:ro_te_vr_dependence_on_Omega}
\vspace{1em}
\end{figure}

Panels {\bf a} and {\bf b} show the weak dependence of density and temperature on $\Omega_\ast$. We can hardly see any differences between the four lines, especially near the coronal base.
The rotation-rate dependences of stellar wind parameters are shown more clearly in Figure \ref{fig:ro_te_vr_dependence_on_Omega}.
Panel {\bf a} shows the dependence of the electron number density at the coronal-base (where $p = 0.03 {\rm \ dyne \ cm^{-2}}$), $n_{e,{\rm cor}}$, on $\Omega_\ast$. In contrast to the observed relation \citep[dashed line, ][]{Ivano03}, our simulations do not show any correlation.
In Panel {\bf b}, we show the coronal-base temperature ($T_{\rm cor}$, circles) and the maximum temperature ($T_{\rm max}$, diamonds) against $\Omega_\ast$, both of which are inconsistent with the relation $T_{\rm cor} \propto \Omega_\ast^{0.45}$ found by \citet{Fionn18}.

These inconsistencies can be explained in the framework of our model.
The observed scaling relations of $n_e$ and $T$ are deduced from X-ray emission that mostly comes from closed magnetic field loops,
while the wind comes from the open-field region \citep[e.g.][]{Cranm09b}.
Therefore, the difference between our result and the observed trends indicates that the density and temperature scale in different ways for open and closed regions. 
To exemplify this, we also show a power-law fitting between $T_{\rm max}$ and $\Omega_\ast$: $T_{\rm max} \propto \Omega_\ast^{0.18}$, which is closer to the relation by \cite{Fionn18}.

The constant coronal temperature with respect to $\Omega_\ast$ is due to the constant Alfv\'en-wave energy flux transmitted into the corona.
As Alfv\'en waves are the only source of coronal heating in our model, their constant energy flux leads to a constant coronal temperature.
The coronal density is in general determined by the energy balance between radiative cooling and conductive heating \citep{Hamme82,Withb88}.
Because the constant temperature yields constant basal conductive heating, 
the coronal density is also kept fixed with $\Omega_\ast$.
Therefore, the constant coronal temperature and density are attributed to constant Alfv\'en-wave energy flux in the corona, which is discussed in more detail in Section 4.

Panel {\bf c} shows the $\Omega_\ast$-dependence of the wind terminal velocity.
For $\Omega_\ast / \Omega_\odot \lesssim 7$, the wind velocity weakly depends on the rotation rate. However, when $\Omega_\ast / \Omega_\odot \gtrsim 7$, the wind velocity drastically increases with rotation.
Beyond a critical rotation rate, the magneto-centrifugal force dominates the force balance in the wind acceleration, resulting in a strong acceleration of the wind \citep{Belch76}.
The critical point $\Omega_\ast / \Omega_\odot \approx 7$ turns out to be the regime-changing point in terms of energy budget.
We will discuss this further in Section \ref{sec:result_energetics}.

\subsection{Angular momentum loss rate (torque)} \label{sec:result_torque}

One of the principal purposes of this work is to investigate whether the Alfv\'en-wave driven magnetic rotator wind model can explain the observed spin-down of low-mass stars.
Ignoring the core-envelope decoupling and internal-structure evolution, the stellar rotational evolution is described as
\begin{align}
  I_\ast \frac{d\Omega_\ast}{dt} = - \tau_w,
\end{align}
where $I_\ast$ is the momentum of inertia of the star.
If one assumes that the wind torque is approximated by $\tau_w \propto \Omega_\ast^{p+1}$, the solution of the rotational evolution yields
\begin{align}
  \Omega_\ast \propto t^{-1/p},
\end{align}
from which the Skumanich relation is reproduced when $p=2$.

In the quasi-steady state, an analytical formulation of torque can be obtained \citep[see e.g.][]{Lamer99}.
The time-averaged mass conservation and magnetic-flux conservation are given by
\begin{align}
	\dot{M}_w &= 4 \pi r^2 f^{\rm open} \rho v_r  = {\rm const.}, \\
	\Phi_{\rm open} &= 4 \pi r^2 f^{\rm open} B_r = {\rm const.}. 
\end{align}
Combining these equations gives the following identity,
\begin{align}
	\frac{\Phi_{\rm open}^2}{16 \pi^2 \dot{M}_w} = r^2 f^{\rm open} \frac{v_A^2}{v_r} = {\rm const.} = r_A^2 v_{r,A}, \label{eq:identity_rA2Mdot}
\end{align}
where we assume that the open flux filling factor at the Alfv\'en point is unity.
For simplicity, we assume that $r_A$ is spherically-symmetric.
The torque is then given by
\begin{align}
	\tau_w = \frac{2}{3} \dot{M}_w r_A^2 \Omega_\ast = \frac{2}{3} \frac{\left( B_{r,\ast} f^{\rm open}_\ast \right)^2}{v_{r,A}} R_\ast^4 \Omega_\ast. \label{eq:torque_scaling1}
\end{align}
where $v_{r,A}$ is the wind velocity at the Alfv\'en point.
Substituting $B_{r,\ast} = 1300 {\rm \ G}$, $R_\ast = 6.96 \times 10^{10} {\rm \ cm}$ and Eq. (\ref{eq:fopen_thiswork}), a semi-analytical expression of $\tau_w$ is obtained:
\begin{align}
  \tau_w = 1.22 \times 10^{30} \left( \frac{v_{r,A}}{v_{g,\odot}} \right)^{-1} \left( \frac{\Omega_\ast}{\Omega_\odot} \right)^{3.4} {\rm \ erg}, \label{eq:torque_analytical}
\end{align}
where $v_{g,\odot} = \sqrt{2GM_\odot/R_\odot} \approx 617 {\rm \ km \ s^{-1}}$ is the escape velocity at the solar surface.
To express $v_{r,A}/v_{g,\odot}$ as a function of $\Omega_\ast$, numerical simulation is required.

\begin{figure}[t]
\centering
\includegraphics[width=70mm]{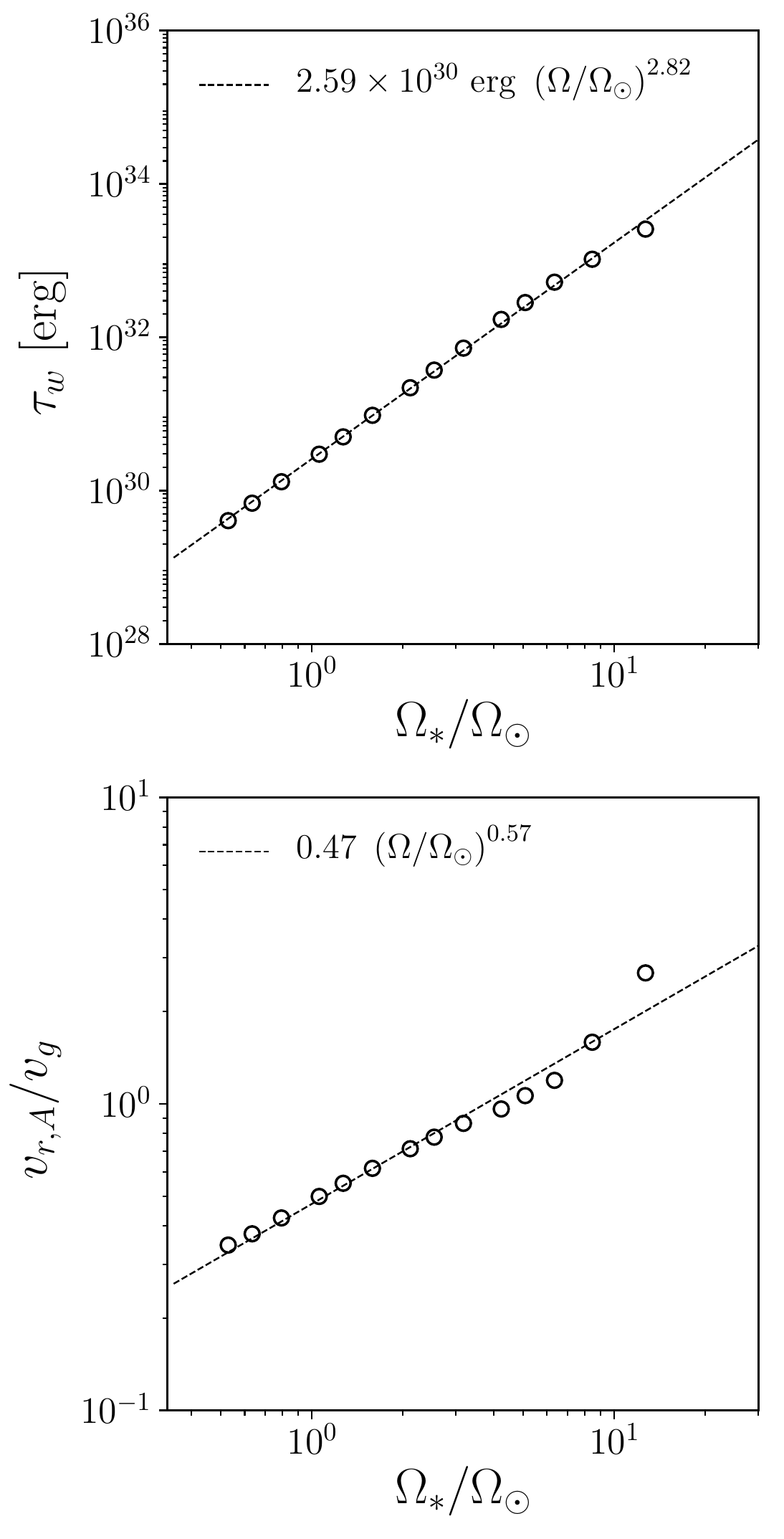}
\vspace{1em}
\caption
{
  Circles in each panel shows the simulated
  {\bf a}: angular momentum loss rates (torques) of the winds, $\tau_w$,
  {\bf b}: stellar wind velocities at the Alfv\'en point, $v_{r,A}$, normalized by the escape velocity at the surface $v_{g,\odot}$.
  Also shown by dashed lines are power-law fittings to circles.
}     
 \label{fig:torque_vrA_versus_Omega}
\vspace{1em}
\end{figure}

The torques calculated from our numerical simulations are shown in Figure \ref{fig:torque_vrA_versus_Omega}.
Panel {\bf a} shows how the wind torque varies with the stellar rotation rate.
In the whole range of $\Omega_\ast$, the simulation results are well fitted by a single power law of
\begin{align}
  \tau_w = 2.59 \times 10^{30} \left( \frac{\Omega_\ast}{\Omega_\odot} \right)^{2.82} {\rm \ erg}, \label{eq:torque_omega}
\end{align}
which yields
\begin{align}
  \Omega_\ast \propto t^{-0.549}.
\end{align}
We note that the data points in the fast-rotator regime deviate slightly from the fit line, which is possibly a result of the regime change (see Section \ref{sec:result_energetics}).
This spin-down law is consistent with the recent gyrochronology relation from \citet{Angus15} who find $P_{\rm rot} \propto t^{0.55}$ .
For $\Omega/\Omega_\odot = 1$, the calculated angular momentum loss rate ($\tau_{w,\odot} = 2.59 \times 10^{30} {\rm \ erg}$) matches with the observed solar-wind torque \citep{Finle18b,Finle19b}.
However, this value is still smaller than the stellar-observation-based empirical value \citep[$\tau_{w,\odot} = 6.3 \times 10^{30} {\rm \ erg}$, ][]{Matt015} by a factor of $2.4$.
One possibility for this gap is that the Sun has smaller amount of open magnetic flux than typical Sun-like stars.
We will discuss this point in more detail in Section \ref{sec:discussion}.

Comparing Eq. (\ref{eq:torque_analytical}) and Eq. (\ref{eq:torque_omega}), one can tell that $v_{r,A}/v_{g,\odot}$ should depend on $\Omega_\ast$, specifically $v_{r,A}/v_{g,\odot} \propto \Omega_\ast^{0.58}$.
This is directly confirmed in Panel {\bf b}.
If we simply assume that $v_{r,A}/v_{g,\odot} \approx 1$, which is not a bad approximation, the torque scales as $\tau_w \propto \Omega^{3.4}$, yielding slower spin-down than the observed one: $\Omega_\ast \propto t^{-0.417}$.
In this respect, the $\Omega_\ast$-dependence of $v_{r,A}/v_{g,\odot}$ is also important in evaluating the spin-down law.

\begin{figure}[t]
\centering
\includegraphics[width=80mm]{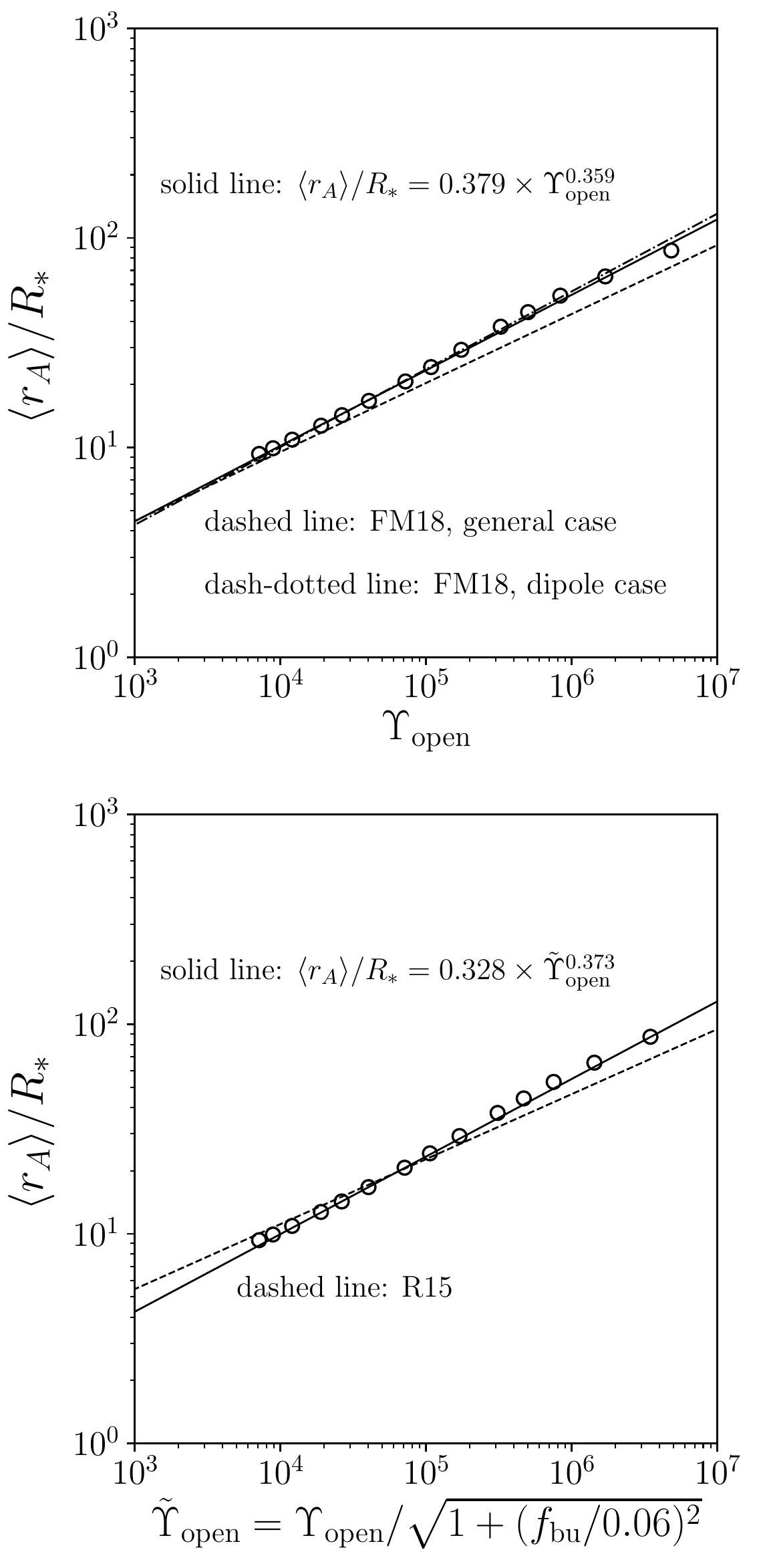}
 \caption{
 	Comparison of Alfv\'en-radius scaling laws from \citet{Finle18a} (top) and \citet{Revil15} (bottom).
 }     
 \label{fig:rA_comparison_with_previous_works}
\vspace{1em}
\end{figure}

\subsection{Alfv\'en radius}
Several works have produced semi-analytic scaling relations for $r_A$ \citep{Kawal88,Matt008,Matt012}.
Here we compare two relations which are based on the open magnetic flux in the wind. These scaling relations are given in terms of the dimensionless wind-magnetization parameter $\Upsilon_{\rm open}$, which is:
\begin{align}
  \Upsilon_{\rm open} &= \frac{\Phi_{\rm open}^2}{R_\ast^2 \dot{M}_w v_{g,\odot}}. 
\end{align}
For comparison, we convert the equatorial Alfv\'en radius from our simulations, $r_A$, to a latitudinally averaged value, $\langle r_A \rangle$, based on the following formulation \citep{Washi93}:
\begin{align}
	\tau_w = \frac{2}{3} \dot{M}_w r_A^2 \Omega_\ast = \dot{M}_w \langle r_A \rangle^2 \Omega_\ast.
\end{align}
Therefore, $\langle r_A \rangle = \sqrt{2/3} \ r_A$, where spherical symmetry has been assumed.

We compare first to the scaling law given by \citet{Finle18a}:
\begin{align}
  \langle r_A  \rangle_{\rm FM18} / R_\ast  =
  \begin{cases}
    0.33 \ {\Upsilon_{\rm open}}^{0.371} \ \ \ \ ({\rm dipole}),  \\
    0.46 \ {\Upsilon_{\rm open}}^{0.329} \ \ \ \ ({\rm general}), 
  \end{cases}
  \label{eq:rA_scaling_Finley}
\end{align}
where the first case is fitted from simulations with only dipole fields and the second case corresponds to a fit using a range of simulations with combinations of dipole, quadrupole and octupole geometries.

The second scaling relation is given by \citet{Revil15} as follows:
\begin{align}
  \langle r_A  \rangle_{\rm R15} / R_\ast  &= 0.64 \left[ \frac{\Upsilon_{\rm open}}{\sqrt{1+ (f_{\rm bu}/0.06)^2}} \right]^{0.31},
  \label{eq:rA_scaling_Reville}
\end{align}
where $f_{\rm bu} = \Omega_\ast R_\ast^{3/2} \left( GM_\ast \right)^{-1/2}$ is the break-up fraction of the rotation speed.

In Figure \ref{fig:rA_comparison_with_previous_works} we compare our results
with \citet{Finle18a} (top panel, dashed and dash-dotted lines) and \citet{Revil15} (bottom panel, dashed line).
In each panel, our results and a fitted power-law are indicated by circles and the solid line.
Both \citet{Revil15} and \citet{Finle18a} are consistent with our results,
indicating that scaling relations for the Alfv\'en radius are robust regardless of simulation setting.
Note that the open flux is an output of the simulations in \citet{Revil15} and \citet{Finle18a} 
while the mass loss rate is mostly controlled by the coronal density and temperature imposed at the boundary condition.
On the other hand, in our calculations the mass-loss rate is an output while the open flux is an input.
A more self-consistent treatment requires a full-sphere simulation
with physics-based coronal heating and chromospheric evaporation.

\begin{figure}[t]
\centering
\includegraphics[width=80mm]{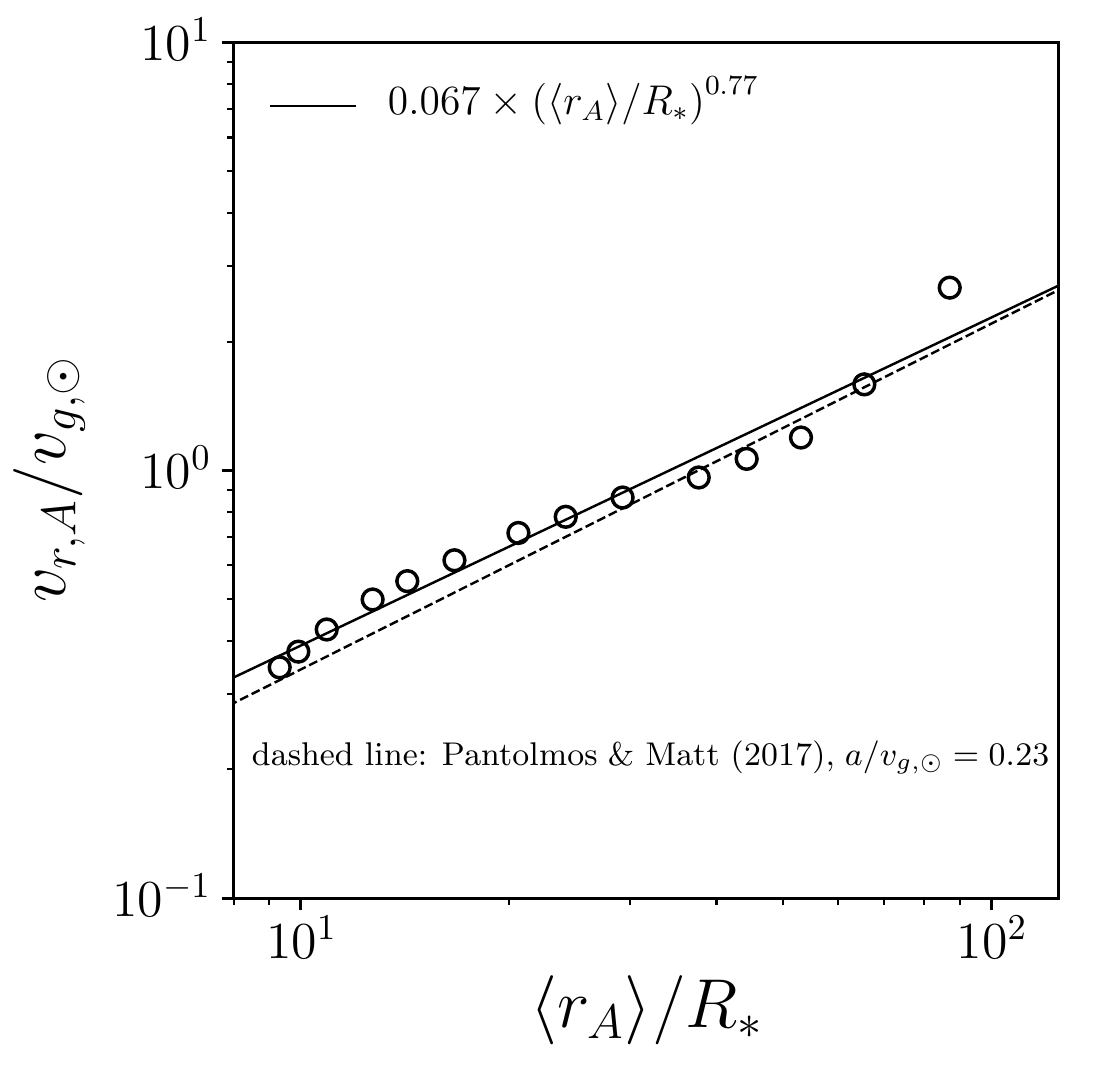}
 \caption{
 	 $\langle r_A \rangle / R_\ast$ versus $v_{r,A}/v_{g,\odot}$ (circles) and a power-law fit
	 ($v_{r,A}/v_{g,\odot} \propto \left( \langle r_A \rangle / R_\ast \right)^{0.77}$, solid line).
	 Also shown by a dashed line is the relation given by \citet{Panto17} with $a/v_{g,\odot}=0.23$.
 }     
 \label{fig:rA_versus_vrA}
\vspace{1em}
\end{figure}
As shown in Figure \ref{fig:rA_comparison_with_previous_works}, our simulations yield a power-law relation between $\langle r_A \rangle / R_\ast$ and $\Upsilon_{\rm open}$ of
\begin{align}
	\langle r_A \rangle / R_\ast  \propto \Upsilon_{\rm open}^{0.36}.
\end{align}
The origin of the exponent $0.36$ is explained as follows.
Rewriting Eq. (\ref{eq:identity_rA2Mdot}) in terms of $\Upsilon_{\rm open}$, one obtains
\begin{align}
	\langle r_A \rangle^2 / R_\ast^2 \propto \Upsilon_{\rm open} \left( \frac{v_{r,A}}{v_{g,\odot}} \right)^{-1}. 
\end{align}
Suppose a power-law relation $v_{r,A} / v_{g,\odot} \propto \left( r_A / R_\ast \right)^q$ is satisfied, then
\begin{align}
	\langle r_A \rangle / R_\ast \propto \Upsilon_{\rm open}^{1/(2+q)}. \label{eq:rA_Upsilon_qvalue}
\end{align}
It is evident from Figure \ref{fig:rA_versus_vrA} that the power-law relation $v_{r,A} / v_{g,\odot} \propto \left( r_A / R_\ast \right)^q$ is satisfied with $q = 0.77$, which yields the exponent in Eq. (\ref{eq:rA_Upsilon_qvalue}) of $1/(2+q) = 0.361$.
Also, our results are consistent with the scaling law by \citet{Panto17} if we adopt the sound-to-escape velocity ratio $a/v_{g,\odot} = 0.23$.
According to \citet{Panto17}, the $q$ value is sensitive to the coronal temperature.
In our model, the coronal temperature is almost constant with respect to $\Omega_\ast$, and thus all our simulations are fitted by a unique $q$ value.

\begin{figure}[t]
\centering
\includegraphics[width=80mm]{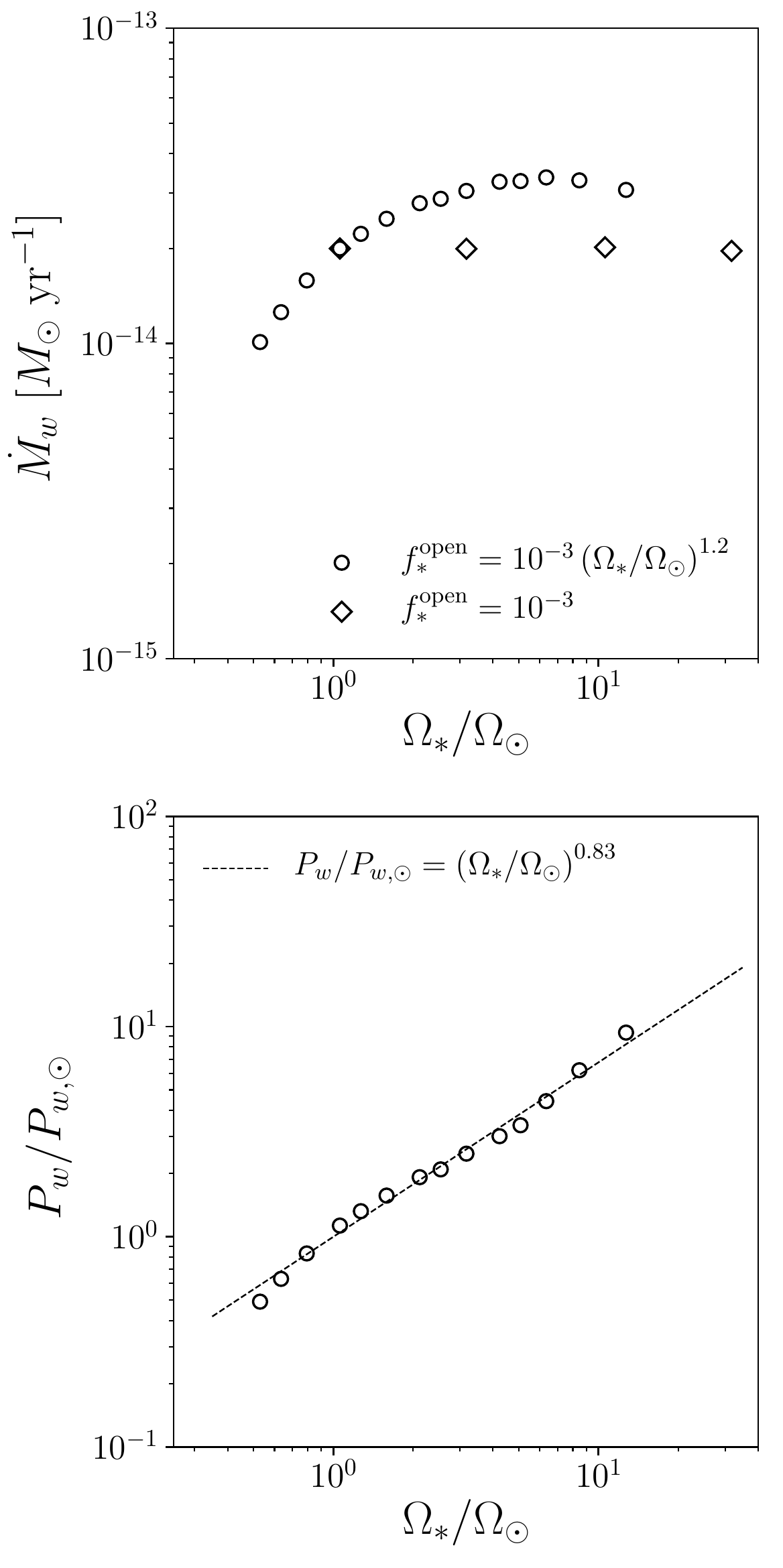}
\caption
{
  (Top) Mass-loss rate, $\dot{M}_w$, versus rotation rate, $\Omega_\ast$.
  Circles show the fiducial cases ($f^{\rm open}_\ast = 10^{-3} \left( \Omega_\ast / \Omega_\odot \right)^{1.2}$) and diamonds show the results with fixed $f^{\rm open}_\ast$ ($f^{\rm open}_\ast = 10^{-3}$, see also Figure \ref{fig:energetics_rot_dependence} and discussions there). 
  (Bottom) Characteristic wind ram pressure, $P_w = 4 \pi r^2 f \rho v_r^2$, normalized by the solar value versus rotation rate, $\Omega_\ast$.
  Shown by dashed line is a power-law fit to the numerical results: $P_w / P_{w,\odot} = \left( \Omega_\ast / \Omega_\odot \right)^{0.83}$. 
 }     
 \label{fig:Mdot_Pwind_versus_Omega}
\vspace{1em}
\end{figure}

\subsection{Mass-loss rate}
In this section, we discuss another interesting topic: the rotation dependence of mass-loss rates, $\dot{M}_w$.
The top panel of Figure \ref{fig:Mdot_Pwind_versus_Omega} shows how $\dot{M}_w$ varies with the stellar rotation rate (circles).
Also shown by diamonds are the results with a fixed open-flux filling factor $f^{\rm open}_\ast = 10^{-3}$ (see also Figure \ref{fig:energetics_rot_dependence}).
$\dot{M}_w$ increases with $\Omega_\ast$ in the slow rotator regime
and saturates around $\dot{M}_w \sim 3.4 \times 10^{-14} \ M_\odot \ {\rm yr^{-1}}$ in the faster rotation cases.
Observations of asterospheric line absorption show that the mass-loss rate tends to increase with X-ray flux \citep{Wood002,Wood005,Wood014}, and thus with rotation rate \citep{Gudel97,Ribas05,Wrigh11,Magau20}.
However, we need to note that
what is actually obtained by the asterospheric observation is the characteristic ram pressure, $P_w = 4 \pi r_{\rm out}^2 \rho_{\rm out} v_{r,{\rm out}}^2$, \citep{Holzw07}
not the mass-loss rate, $\dot{M}_w = 4 \pi r_{\rm out}^2 \rho_{\rm out} v_{r,{\rm out}}$.
Bearing this in mind, we henceforth focus on $P_w$ for comparison with observation.

\begin{figure}[t]
\centering
\includegraphics[width=80mm]{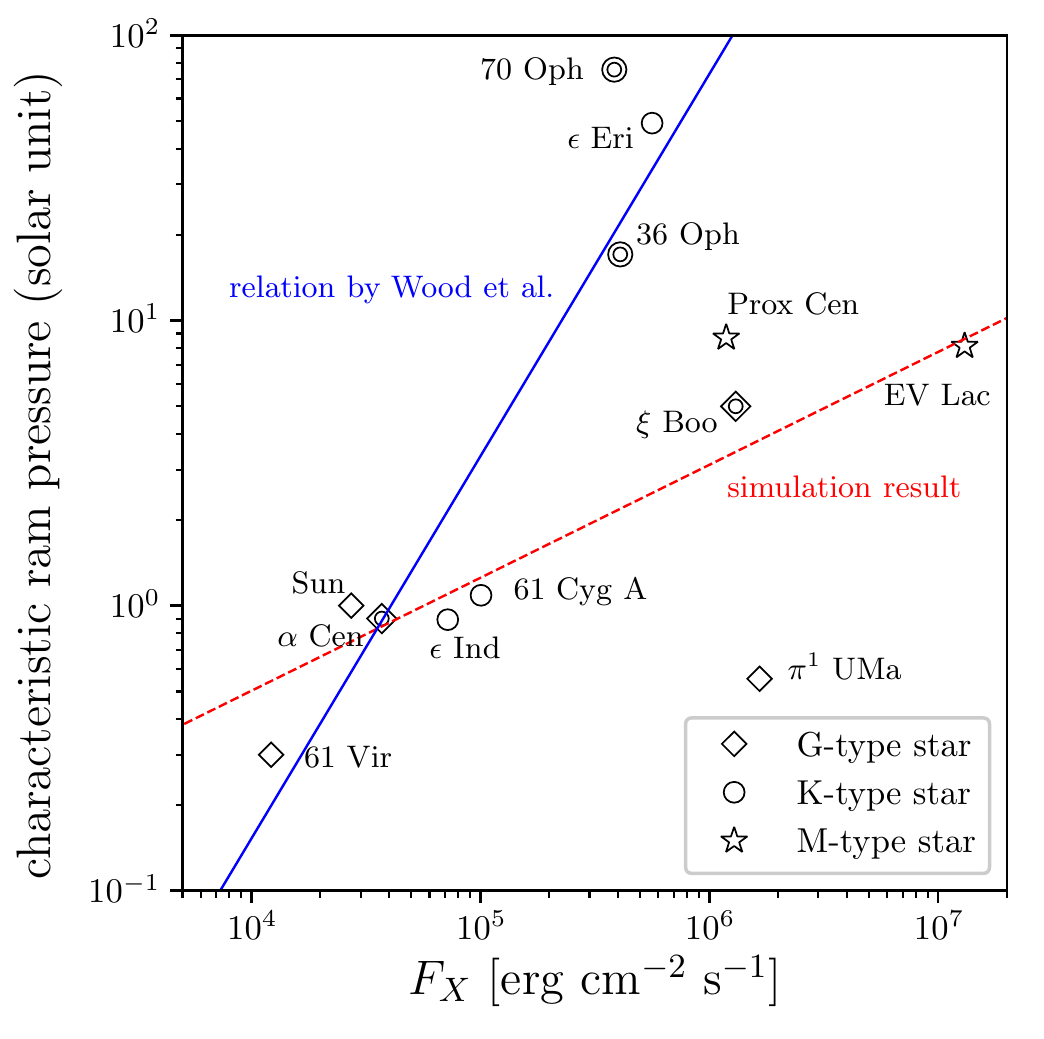}
\caption
{
  X-ray flux $F_X$ versus characteristic ram pressure normalized by solar value $P_w / P_{w,\odot}$.
  Symbols indicate the asterospheric observations; diamonds, circles and stars stand for G-type, K-type and M-type stars, respectively, and binaries are doubly marked with corresponding symbols../a
  Also shown by red dashed and blue solid lines show our simulation result and the empirical relation proposed by \citet{Wood005}, respectively.
 }     
 \label{fig:Fx_versus_Mdot}
\vspace{1em}
\end{figure}

The bottom panel of Figure \ref{fig:Mdot_Pwind_versus_Omega} shows the relation between $P_w / P_{w,\odot}$ and $\Omega_\ast / \Omega_\odot$, which has a power-law relation of
\begin{align}
  P_w / P_{w,\odot} = \left( \Omega_\ast / \Omega_\odot \right)^{0.83}, \label{eq:ram_pressure_Omega}
\end{align}
where we set $P_{w,\odot} = 5.0 \times 10^{19} {\rm \ dyne}$.
Following \citet{Wrigh11}, we convert the rotation rate $\Omega_\ast$ to the X-ray flux $F_X$ as
\begin{align}
  F_X / F_{X,\odot} = \left( \Omega_\ast / \Omega_\odot \right)^{2.18}, \label{eq:Fx_Omega}
\end{align}
where $F_{X,\odot} = 3 \times 10^4 {\rm \ erg \ cm^{-2} \ s^{-1}}$.
Note that 1. all of our simulation runs are in the unsaturated regime (in which stellar activities correlate with stellar rotation) and 2. there is a one to one relation between $F_X$ and $L_X/L_{\rm bol}$ because the stellar radius and luminosity are fixed in our simulations.
Combining Eq.s (\ref{eq:ram_pressure_Omega}) and (\ref{eq:Fx_Omega}),
\begin{align}
  P_w / P_{w,\odot} = \left( F_X / F_{X,\odot} \right)^{0.38}. \label{eq:Fx_ram_pressure}
\end{align}
Figure \ref{fig:Fx_versus_Mdot} shows $F_X$-$P_w$ for the asterospheric
observations taken from \citet{Wood014} (symbols), the empirical relation from \citet{Wood005} (blue solid line), and our result (Eq. (\ref{eq:Fx_ram_pressure}), red dashed line).
Our simulation result is consistent (within a factor 3) with the observations of
61 Vir, Sun, $\alpha$ Cen, $\varepsilon$ Ind, 61 Cyg A, $\xi$ Boo, Prox Cen, and EV Lac.
A similar trend is found in the work of \citet{Holzw07}.
Our model is able to explain a good fraction of the observations, although there exists non-negligible offsets for three K-dwarfs (36 Oph, 70 Oph, $\varepsilon$ Eri).
It is left for future work to test whether Eq. (\ref{eq:ram_pressure_Omega}) and Eq. (\ref{eq:Fx_ram_pressure}) are valid for non-Sun-like stars.

\section{Wind energetics}
\label{sec:result_energetics}

The physics of the stellar wind heating and acceleration can be inferred by following the energy flow from the stellar surface to interplanetary space.
For example, one can estimate the stellar wind mass-loss rate analytically, based on wind energetics \citep{Hanst95,Cranm11,Suzuk18}.
To understand what causes the saturation of mass-loss rate, the energy budget in the stellar wind is discussed.

\subsection{Energy conservation}
\label{sec:analytical_energy_cons}
After time averaging, the energy conservation law is written as follows:
\begin{align}
  \frac{d}{dr} \left( L_K + L_E + L_A - L_C - L_G \right)  =  - 4 \pi r^2 f^{\rm open} Q_{\rm rad},
  \label{eq:steady_energy_conservation}
\end{align}
where
\begin{subequations}
\begin{align}
  L_K &= \frac{1}{2} \rho v_r^3 4 \pi r^2 f^{\rm open}, \\
  L_E &= \frac{\gamma}{\gamma-1} p v_r 4 \pi r^2 f^{\rm open}, \\
  L_A &= \left[ \left( \frac{1}{2} \rho \vec{v}_\perp^2 + \frac{\vec{B}_\perp^2}{4 \pi} \right) v_r - \frac{B_r}{4 \pi} \left( \vec{v}_\perp \cdot \vec{B}_\perp \right) \right] 4 \pi r^2 f^{\rm open}, \\
  L_C &= - F_C 4 \pi r^2 f^{\rm open}, \\
  L_G &=  \rho v_r \frac{G M_\ast}{r} 4 \pi r^2 f^{\rm open} =  \dot{M}_w \frac{G M_\ast}{r}.
\end{align}
\end{subequations}
$L_A$, $L_E$, $L_A$, $L_C$, and $L_G$ correspond to the wind kinetic energy flux, enthalpy flux, Alfvén-wave energy flux, conductive flux, and gravitational energy flux, respectively.
Bearing in mind that $\theta$ and $\phi$ components stand for Alfv\'en waves and rotation, 
we can decompose $L_A$ as
\begin{subequations}
\begin{align}
  L_A &= L_A^{\rm wav} + L_A^{\rm rot}, \\
  L_A^{\rm wav} &= \left[ \left( \frac{1}{2} \rho v_\theta^2 + \frac{B_\theta^2}{4 \pi} \right) v_r - \frac{B_r}{4 \pi} v_\theta B_\theta \right] 4 \pi r^2 f^{\rm open}, \\
  L_A^{\rm rot} &= \left[ \left( \frac{1}{2} \rho v_\phi^2 + \frac{B_\phi^2}{4 \pi} \right) v_r - \frac{B_r}{4 \pi} v_\phi B_\phi \right] 4 \pi r^2 f^{\rm open},
\end{align}
\end{subequations}
where $L_A^{\rm wav}$ and $L_A^{\rm rot}$ correspond to the luminosities of Alfv\'en waves and magneto-rotation.
%Note that the luminosities are measured in the radial direction, not in the mean magnetic field direction.

\subsection{Energetics in the corona and above}\label{sec:energetics_corona_above}

The energy conservation is more simply approximated above the coronal base, where the enthalpy flux and the radiative loss is negligibly small.
\begin{align}
  \frac{d}{dr} \left( L_K + L^{\rm wav}_A + L^{\rm rot}_A - L_C - L_G \right) \approx 0. \label{eq:energy_conservation_simplified}
\end{align}
Integrating this equation from the coronal base to the distant stellar wind and ignoring minor components, one obtains
\begin{align}
   L^{\rm wav}_{A,{\rm cor}} +  L^{\rm rot}_{A,{\rm cor}} - L_{C,{\rm cor}} - L_{G,{\rm cor}} \approx L_{K,{\rm out}} + L^{\rm rot}_{A,{\rm out}}, \label{eq:luminosity_cons_approximated}
\end{align}
where $X_{\rm cor}$ and $X_{\rm out}$ denote $X$ measured at the coronal base and outer boundary, respectively.

\begin{figure}[t]
\centering
\includegraphics[width=80mm]{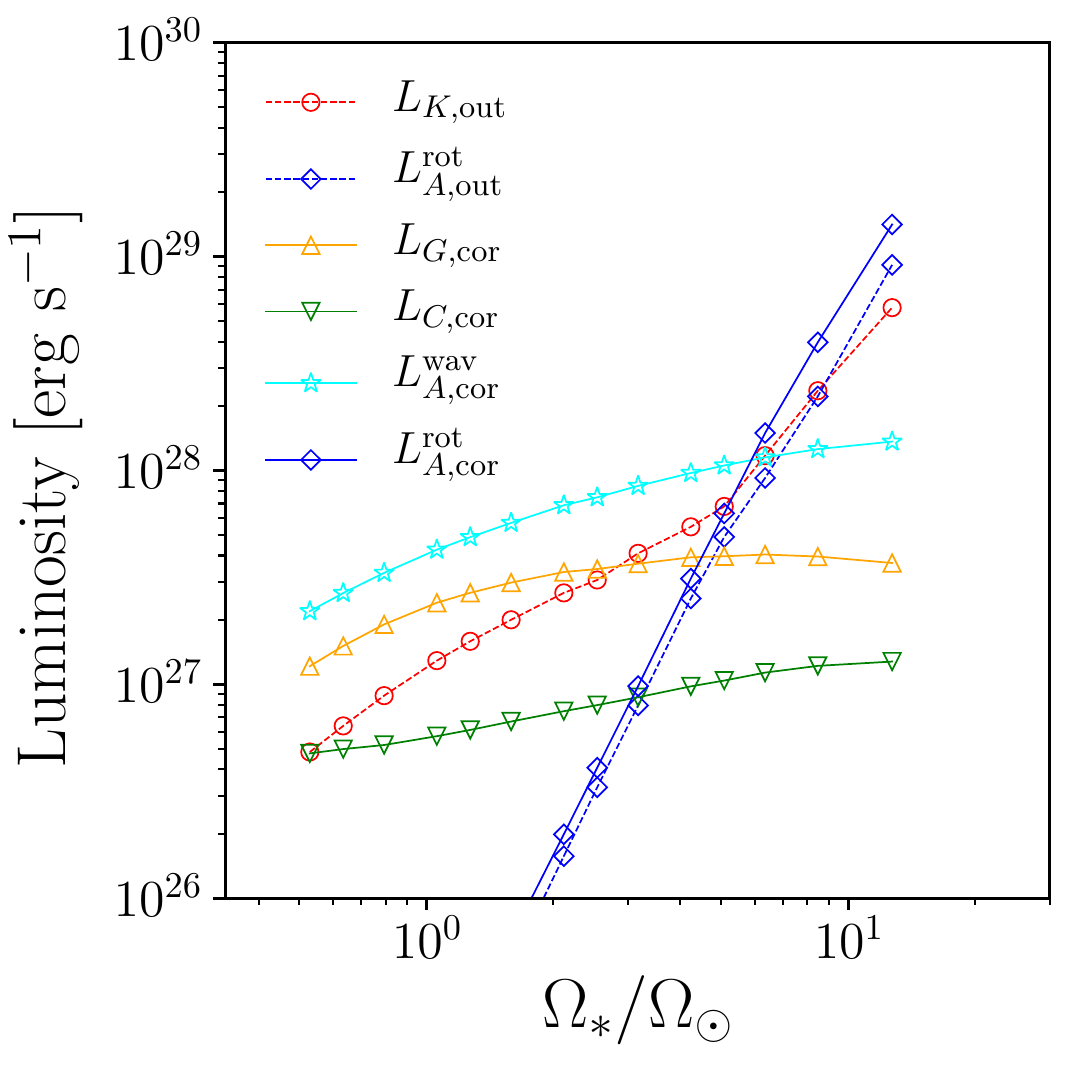}
\caption
{
  Luminosities in Eq. (\ref{eq:luminosity_cons_approximated}) as functions of rotation rate $\Omega_\ast$.
  Shown are the 
   wind kinetic energy flux at the outer boundary $L_{K,{\rm out}}$ (red-dashed line),
  rotational energy flux at the outer boundary $L_{A,{\rm out}}^{\rm rot}$ (blue-dashed line) ,
   gravitational energy flux at the coronal base $L_{G,{\rm cor}}$ (orange-solid line),
   downward conductive energy flux at the coronal base $L_{C,{\rm cor}}$ (green-solid line),
   Alfv\'en-wave energy flux at the coronal base $L_{A,{\rm cor}}^{\rm wav}$ (cyan-solid line),
  and rotational energy flux at the coronal base $L_{A,{\rm cor}}^{\rm rot}$ (blue-solid line),
  respectively.
 }     
 \label{fig:energetics_vs_Omega}
\vspace{1em}
\end{figure}

Taking the coronal base as $r = 1.02 R_\ast$, we have confirmed that this approximated energy balance relation is satisfied to within $2\%$ error.
In Figure \ref{fig:energetics_vs_Omega}, we plot each term in Eq. (\ref{eq:luminosity_cons_approximated})
as a function of rotation rate $\Omega_\ast$.
%Each line indicates $L_{K,{\rm out}}$ (red-dashed line), $L_{A,{\rm out}}^{\rm rot}$ (blue-dashed line), $L_{G,{\rm cor}}$ (orange-solid line),
%$L_{C,{\rm cor}}$ (green-solid line), $L^{\rm wav}_{A,{\rm cor}}$ (cyan-solid line), and $L^{\rm rot}_{A,{\rm cor}}$ (blue-solid line), respectively.
Figure \ref{fig:energetics_vs_Omega} has several features:
\begin{enumerate}
\item In the slow-rotator regime ($\Omega_\ast / \Omega_\odot \lesssim 4$),
  the dominant coronal energy injection is by Alfv\'en waves: $L^{\rm wav}_{A,{\rm cor}} \gg L^{\rm rot}_{A,{\rm cor}}$.
  In this regime, the energy balance is approximated as $L^{\rm wav}_{A,{\rm cor}} \approx L_{G,{\rm cor}} + L_{K,{\rm out}}$, as assumed by \citet{Cranm11}.
\item In the fast-rotator regime ($\Omega_\ast / \Omega_\odot \gtrsim 10$),
  the rotation components becomes dominant: $L^{\rm wav}_{A,{\rm cor}} \ll L^{\rm rot}_{A,{\rm cor}}$.
  The energy balance relation is then $L^{\rm rot}_{A,{\rm cor}} \approx L_{K,{\rm out}} + L^{\rm rot}_{A,{\rm out}}$.
\item The ``regime change'' from wave-driven wind
  ($L^{\rm wav}_{A,{\rm cor}} > L^{\rm rot}_{A,{\rm cor}}$) to rotation-driven wind
  ($L^{\rm wav}_{A,{\rm cor}} < L^{\rm rot}_{A,{\rm cor}}$) takes place around $\Omega_\ast / \Omega_\odot \approx 7$,
  or equivalently $P_{\rm rot}\approx 3.6 {\rm \ day}$.
  Note that the regime-changing period strongly depends on the filling factor of open flux.
\end{enumerate}

\begin{figure}[t]
\centering
\includegraphics[width=80mm]{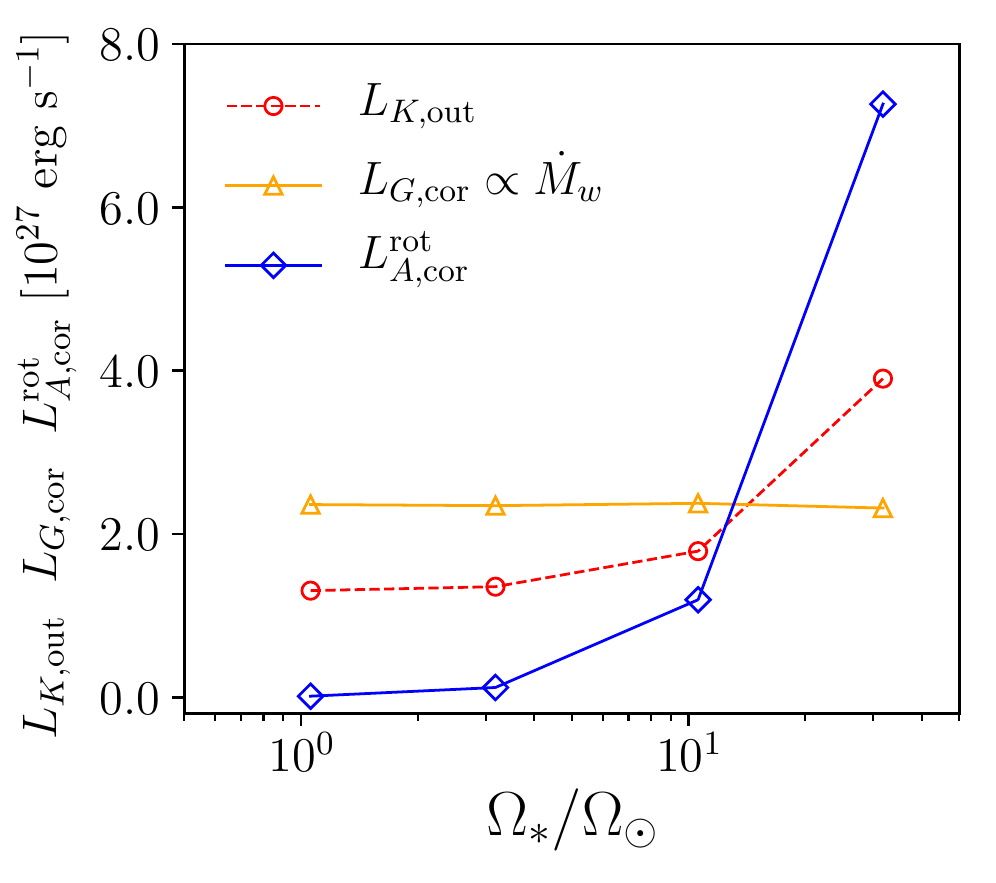}
\caption
{
  Energy-flux dependence on the rotation rate $\Omega_\ast$ the open-flux filling factor fixed to $f^{\rm open}_\ast=10^{-3}$.
  Definitions of the lines and markers are the same as Figure \ref{fig:energetics_vs_Omega}.
 }     
 \label{fig:energetics_rot_dependence}
\vspace{1em}
\end{figure}

An interesting behavior of the rotation-driven wind is that,
in spite of the rapid increase of rotational energy injection $L_{A,{\rm cor}}^{\rm rot}$ with $\Omega_\ast$,
the $L_{G,{\rm cor}} \propto \dot{M}_w$ does not increase.
This is because the wind density is determined by the energy injected below the sonic (slow-magnetosoic) point \citep{Hamme82,Leer082,Hanst95,Hanst12}.
The magneto-rotational acceleration (magneto-centrifugal force) works in the super-sonic region and works to accelerate the stellar wind without increasing mass-loss rate.
To show this, we perform a set of test simulations with different rotation rates ($P_{\rm rot} = 24, 8, 2.4, 0.8 {\rm \ day}$) and fixed open-flux filling factor to $f^{\rm open}_\ast = 10^{-3}$ to see the purely rotational effect on the wind.
Figure \ref{fig:energetics_rot_dependence} shows $L_{K,{\rm out}}$ (black), $L_{G,{\rm cor}} \propto \dot{M}_w$ (blue) and $L_{A,{\rm cor}}^{\rm rot}$ (red)
as functions of $\Omega_\ast$.
The gravitational luminosity $L_{G,{\rm cor}}$ (or the mass-loss rate) does not respond to the increase of rotational energy flux.
Instead, the terminal wind kinetic energy flux increases with rotational energy.
For this reason, enhanced rotation rate does not lead to enhanced mass-loss rate.

\begin{figure}[t]
\centering
\includegraphics[width=80mm]{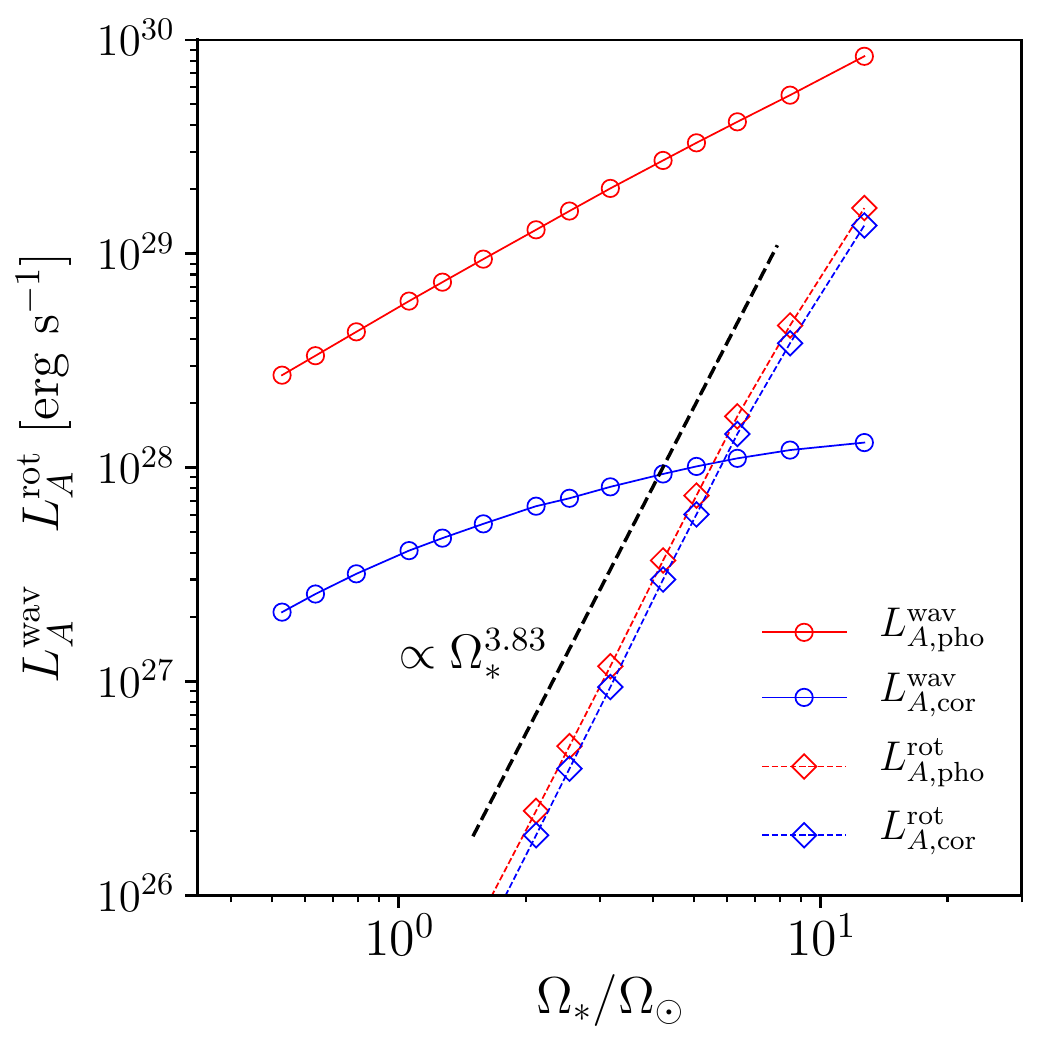}
\caption
{
  $\Omega_\ast$-dependences of Poynting-flux luminosities.
  Alfv\'en-wave luminosities are shown by solid lines for photospheric (red) and coronal-base (blue) values.
  Rotational luminosities are shown by dashed lines for photospheric (red) and coronal-base (blue) values.
  A semi-analytical relation $L_{A,{\rm cor}}^{\rm rot} \propto \Omega_\ast^{3.83}$ is indicated by the dashed line.
 }     
 \label{fig:energy_transmission_vs_Omega}
\vspace{1em}
\end{figure}

\begin{figure*}[t]
\centering
\includegraphics[width=140mm]{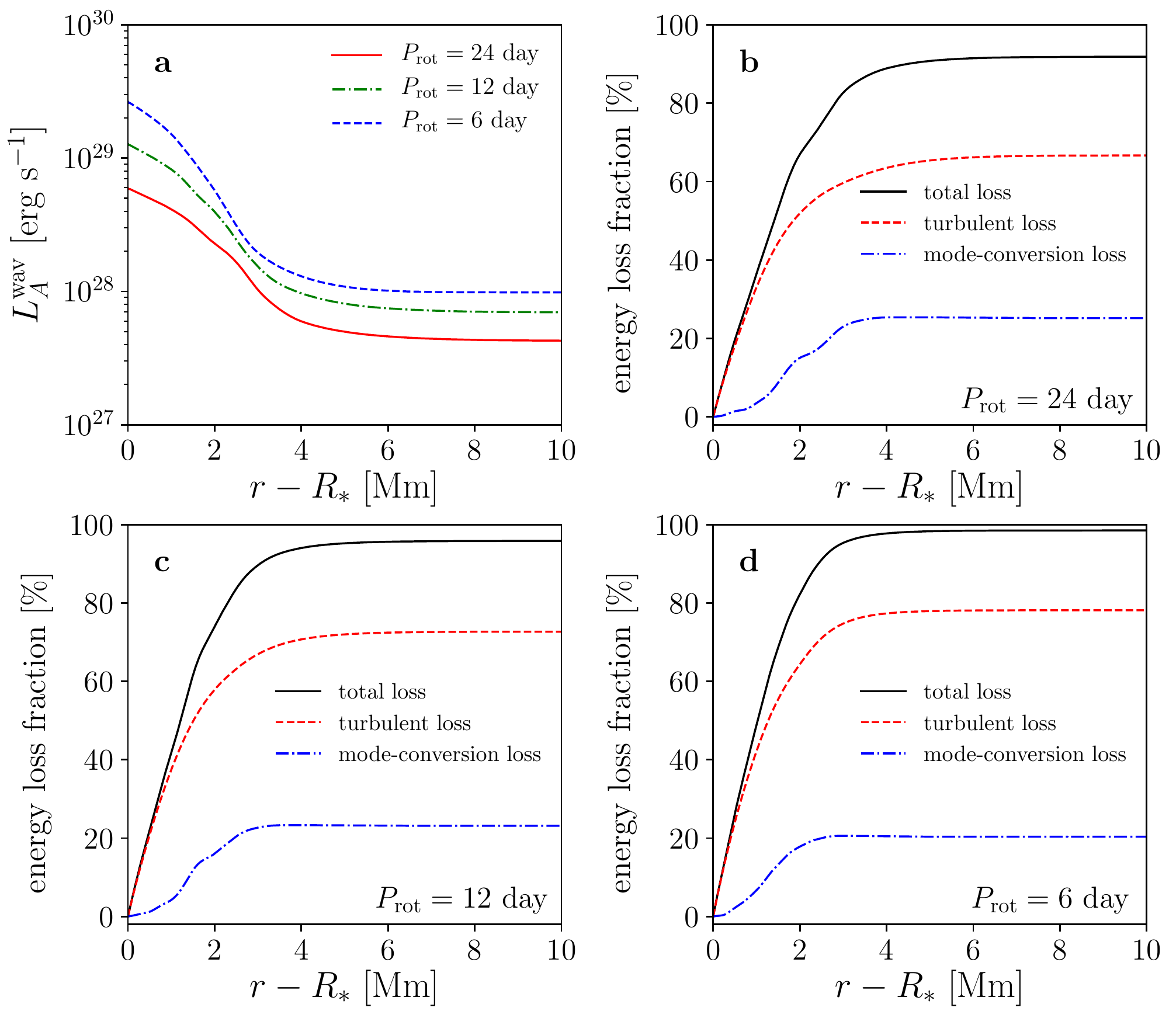}
\caption
{
  Alfv\'en-wave energy loss in the chromosphere.
  {\bf a}: Alfv\'en-wave luminosities versus height for different rotation rates:
  $P_{\rm rot} = 24 {\rm \ day}$ (red-solid line),
  $P_{\rm rot} = 12{\rm \ day}$ (green-dash-dotted line),
  $P_{\rm rot} = 6 {\rm \ day}$ (blue-dashed line).
  {\bf b}: normalized energy-loss fraction $\Delta L_A^{\rm wav} / L_{A,{\rm pho}}^{\rm wav}$ (black-solid line) for $P_{\rm rot} = 24 {\rm \ day}$.
  Energy loss through turbulence and mode-conversion are also shown with a red-dashed line and a blue-dash-dotted line, respectively.
  {\bf c} and {\bf d}: same as Panel {\bf b} now for $P_{\rm rot} = 12 {\rm \ day}$ and $P_{\rm rot} = 6 {\rm \ day}$, respectively.
}     
 \label{fig:wave_luminosity_comparison_prot}
\vspace{1em}
\end{figure*}

An approximated relation for the mass-loss rate is derived from our analysis.
Consider the energy balance assumed in \citet{Cranm11} of
\begin{align}
  L_{K,{\rm out}} + L_{G,{\rm cor}} \approx L_{A,{\rm cor}}^{\rm wav}. 
\end{align}
We have already shown that this holds for the slow-rotator regime but breaks down in the presence of large rotational energy injection.
However Figure \ref{fig:energetics_vs_Omega} shows that, in the whole range of $\Omega_\ast$, $L_{A,{\rm cor}}^{\rm wav}$ is well approximated as
\begin{align}
  L_{A,{\rm cor}}^{\rm wav} \approx 2 L_{G,{\rm cor}}, 
\end{align}
which yields
\begin{align}
  \dot{M}_w \approx \frac{L_{A,{\rm cor}}^{\rm wav}}{v_{g,\odot}^2}.
  \label{eq:massloss_approximation}
\end{align}
Eq. (\ref{eq:massloss_approximation}) is validated as follows.
As we have already shown, the magneto-centrifugal force works to enhance the wind velocity but not to increase the mass-loss rate.
In other words, the mass-loss rate remains unchanged even if the magneto-centrifugal force does not work.
Therefore, the rotational terms in the energy budget equation (\ref{eq:luminosity_cons_approximated}) can be ignored in discussing the mass-loss rate.
Ignoring the conductive flux that are always minor $L_{C,{\rm cor}}$, the energy conservation is reduced to
\begin{align}
  L_{K,{\rm out}} + L_{G,{\rm cor}} = \frac{1}{2} \dot{M}_w \left( v_{r,{\rm out}}^2 + v_{g,\odot}^2 \right) \approx L_{A,{\rm cor}}^{\rm wav}.
  \label{eq:energetics_worotation}
\end{align}
We have confirmed by numerical simulations (not shown here) that, in the absence of magneto-centrifugal force, $v_{r,{\rm out}}$ is always approximated by $v_{g,\odot}$.
With $v_{r,{\rm out}} \approx v_{g,\odot}$, Eq. (\ref{eq:massloss_approximation}) is derived from Eq. (\ref{eq:energetics_worotation}).
Now, it turns out that the saturation of mass-loss rate comes from the saturation of $L_{A,{\rm cor}}^{\rm wav}$ (energy flux of Alfv\'en wave transmitted into the corona),
which is further discussed in the following section.

\subsection{Alfv\'en-wave energetics in the chromosphere} \label{sec:result_alfven_wave}
We further investigate the rotation dependence of Alfv\'en-wave energy flux (or luminosity) at the coronal base $L_{A,{\rm cor}}^{\rm wav}$.
Solid lines in Figure \ref{fig:energy_transmission_vs_Omega} shows the $\Omega_\ast$-dependences of the wave luminosities measured at the photosphere ($L_{A,{\rm pho}}^{\rm wav}$, red line), and the coronal base ($L_{A,{\rm cor}}^{\rm wav}$, blue line).
Dashed lines represent the rotational luminosities: $L_{A,{\rm pho}}^{\rm rot}$ (red) and $L_{A,{\rm cor}}^{\rm rot}$ (blue).
The photospheric value is measured $20 {\rm km}$ above the stellar surface, eliminating the direct influence of lower boundary condition.
We can tell several interesting features from Figure \ref{fig:energy_transmission_vs_Omega}:
\begin{enumerate}
\item Near the stellar surface, the wave energy flux is always larger than the rotation energy flux in the parameter range of our simulations.
  Our fastest rotating case has a surface rotation velocity of $\sim 25 {\rm \ km \ s^{-1}}$,
  which is much larger than the wave amplitude of $\sim 1.2 {\rm \ km \ s^{-1}}$.
  However, because the azimuthal magnetic field $B_\phi$ is  small in the lower atmosphere (as the Weber-Davis solution predicts),
  the energy flux of magneto-rotation remains smaller than the wave energy flux.
\item $L_{A,{\rm cor}}^{\rm wav}$ is much smaller than $L_{A,{\rm pho}}^{\rm wav}$.
  This means that a large fraction of Alfv\'en waves dissipates between the photosphere and coronal base.
  Interestingly, the ``energy transmission rate'', $L_{A,{\rm cor}}^{\rm wav} / L_{A,{\rm pho}}^{\rm wav}$, decreases with $\Omega_\ast$.
  This is why $L_{A,{\rm cor}}^{\rm wav}$ saturates with respect to $\Omega_\ast$ at rapid rotation in spite of the power-law dependence of $L_{A,{\rm pho}}^{\rm wav}$ on $\Omega_\ast$.
\item In contrast to the wave luminosity, the rotational luminosity hardly decreases between the photosphere and coronal base.
  The rotational luminosity steeply increases with $\Omega_\ast$, which is consistent with semi-analytical predictions ($L_{A,{\rm cor}}^{\rm rot} \propto \Omega_\ast^{3.83}$, see Appendix \ref{app:rotational_luminosity}).
  As a result, at the coronal base, the rotational energy flux overtakes the wave energy flux in the fast rotator regime,
  typically from $P_{\rm rot} \lesssim 4 {\rm \ day}$.
  This transition is responssible for the wind regime change discussed in Section \ref{sec:energetics_corona_above}
\end{enumerate}
To summarize, the stellar wind experiences a regime change at $P_{\rm rot} \approx 4 {\rm \ day}$.
This results from a significant decrease in coronal Alfv\'en-wave energy flux which is overtaken by the rotational energy flux at $P_{\rm rot} \approx 4 {\rm \ day}$.
Note that the Alfv\'en-wave energy flux at the stellar surface is always larger than the rotation energy flux.

\begin{figure}[t]
\centering
\includegraphics[width=80mm]{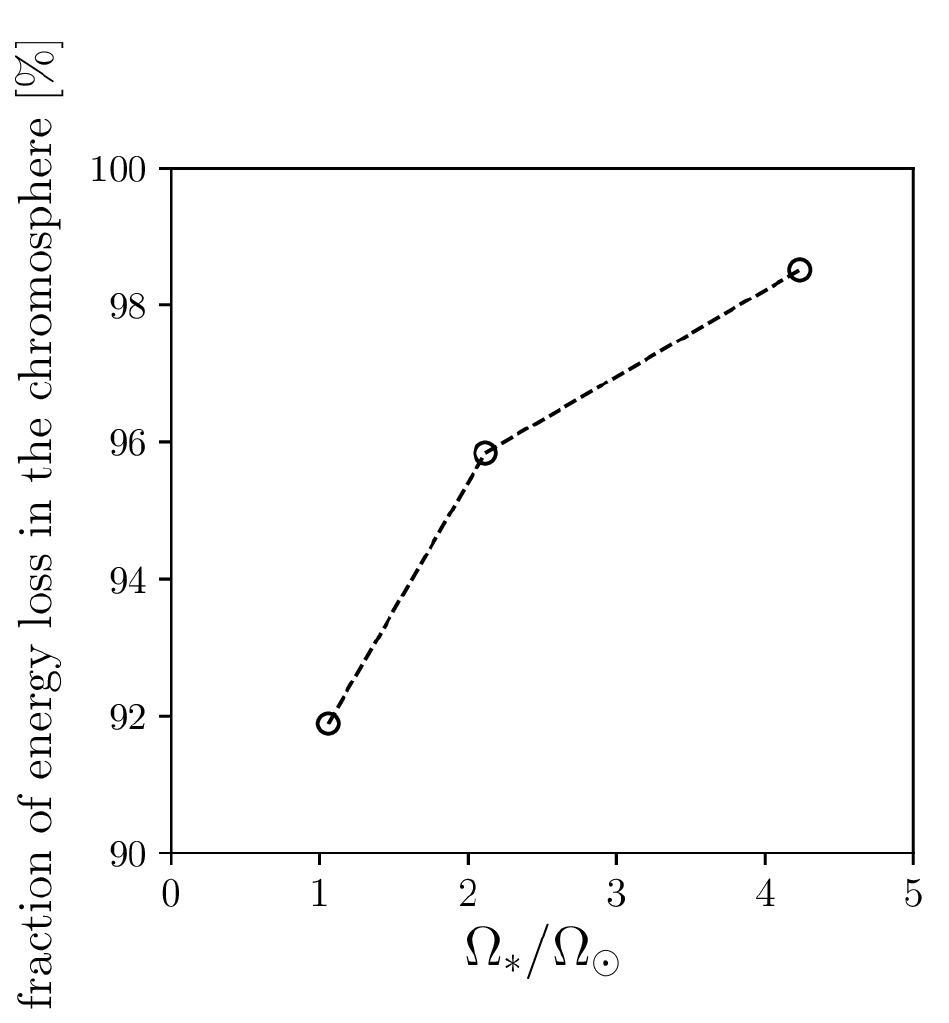}
\caption
{
   Fraction of Alfv\'en-wave energy loss in the chromosphere ($\Delta L_A^{\rm wave}/L^{\rm wav}_{A,{\rm pho}}$ measured at $r - R_\ast = 10 {\rm \ Mm}$) as a function of rotation rate.
 }     
 \label{fig:total_loss_fraction_vs_Omega}
\vspace{1em}
\end{figure}
To reveal the reason for the saturation of Alfv\'en wave luminosity, we analyse the Alfv\'en wave energy loss in the chromosphere.
The conservation of Alfv\'en-wave energy flux (luminosity) is given as follows:
\begin{align}
    &\frac{\partial}{\partial t} \left( \frac{1}{2} \rho v_\theta^2 + \frac{B_\theta^2}{8 \pi} \right) + \frac{1}{4 \pi r^2 f^{\rm open}} \frac{\partial}{\partial r} L_A^{\rm wav}
  = - \varepsilon_{r \leftrightarrow \theta} - Q_{\rm turb},
\end{align}
where $Q_{\rm turb}$ is the turbulent dissipation and $\varepsilon_{r \leftrightarrow \theta}$ represents the energy conversion between
Alfv\'en wave and longitudinal motion:
\begin{align}
  Q_{\rm turb} &= c_d \rho \frac{\left| z_\theta^+ \right| {z_\theta^-}^2 + \left| z_\theta^- \right| {z_\theta^+}^2}{4 \lambda_\perp}, \\
  \varepsilon_{r \leftrightarrow \theta} &= - v_r \frac{\partial}{\partial r} \left( \frac{B_\theta^2}{8 \pi} \right)
  + v_r \left( \rho v_\theta^2 - \frac{B_\theta^2}{4 \pi} \right) \frac{d}{dr} \ln \left( rf^{\rm open} \right).
\end{align}
In this paper, we shall interpret $\varepsilon_{r \leftrightarrow \theta}$ as mode conversion,
since the energy conversion between transverse and longitudinal waves is described by this term.
Note that the mode conversion works efficiently in the chromosphere \citep{Rosen02,Bogda03},
both transforming longitudinal waves to transverse waves \citep{Schun06,Shoda18b} and transverse waves to longitudinal waves \citep{Hollw82a,Kudoh99,Matsu10a}.
We can define the loss fractions as
\begin{align}
  \Delta L_{A,{\rm tot}}^{\rm wav} &= \Delta L_{A,{\rm turb}} + \Delta L_{A,r \leftrightarrow \theta}, \\
  \Delta L_{A,{\rm turb}} &= \int_{R_\ast}^r dr \ 4 \pi r^2 f^{\rm open} Q_{\rm turb} , \\ 
  \Delta L_{A,r \leftrightarrow \theta} &= \int_{R_\ast}^r dr \ 4 \pi r^2 f^{\rm open} \varepsilon_{r \leftrightarrow \theta}.
\end{align}
In Figure \ref{fig:wave_luminosity_comparison_prot}, we show the Alfv\'en-wave energy loss in the chromosphere.
In Panel {\bf a}, we show $L_A^{\rm wav}$ as a function of height for different rotation periods.
In Panels {\bf b}-{\bf d}, the relative fractions of energy loss $\Delta L_A^{\rm wav} / L_{A,{\rm pho}}^{\rm wav}$ are plotted for each case
({\bf b}: $P_{\rm rot} = 24 {\rm \ day}$, {\bf c}: $P_{\rm rot} = 12 {\rm \ day}$, {\bf d}: $P_{\rm rot} = 6 {\rm \ day}$),
where
total loss $\Delta L_{A,{\rm tot}}^{\rm wav} / L_{A,{\rm pho}}^{\rm wav}$,
turbulence loss $\Delta L_{A,{\rm turb}} / L_{A,{\rm pho}}^{\rm wav}$,  
mode-conversion loss  $\Delta L_{A,r \leftrightarrow \theta} / L_{A,{\rm pho}}^{\rm wav}$
are shown by black-solid, red-dashed and blue-dash-dotted lines, respectively.
Black lines in Panels {\bf b}-{\bf d} show that more than $90 \%$ of the Alfv\'en-wave energy flux is lost in the chromosphere.
Moreover, the energy loss fraction is larger for the faster rotating case.
Figure \ref{fig:total_loss_fraction_vs_Omega} shows the trend of the wave-energy loss fraction measured in the corona ($r-R_\odot = 10{\rm \ Mm}$), in which larger loss fraction is clearly seen for faster rotator.
Aforementioned saturation of coronal Alfv\'en-wave energy flux $L_{A,{\rm cor}}^{\rm wav}$ is caused by this enhanced dissipation in fast rotators.

Comparing red lines in Panels {\bf b}-{\bf d},
the increased energy loss is attributed to the increased turbulent loss.
Given that the magnetic filling factor increases with rotation rate,
the increased turbulent loss is a natural consequence for the following two reasons:
\begin{enumerate}
\item When the filling factor is large, magnetic flux expansion is suppressed because one flux tube merges with the adjacent one after small expansion.
  The vortex size of turbulence (correlation length) is expected to expand with the flux tube, and thus remains small for fast rotators that are expected to have large open-flux filling factors.
  Therefore Alfv\'en waves dissipate on smaller timescales, or equivalently, dissipate more quickly.
\item The coronal magnetic field is stronger for cases with faster rotation.
  As a result, Alfv\'en-wave reflection at the transition region is enhanced.
  Therefore Alfv\'en wave turbulence (triggered by the Alfv\'en-wave reflection), should be promoted.
\end{enumerate}

Although the turbulent dissipation plays an important role in our model, 
we expect that our conclusion is not strongly affected by the amount of turbulent dissipation.
Let us consider an extreme case with no turbulent dissipation.
Alfv\'en waves propagate through the chromosphere with less dissipation and are strongly reflected at the transition region \citep{Cranm05,Verdi07,Revil18}.
Reflected Alfv\'en waves propagate backward without turbulent dissipation and reach the stellar surface.
Since the net upward Alfv\'en-wave energy flux is reduced if more downward Alfv\'en waves are present, reduced turbulent dissipation leads to reduced $L_{A,{\rm pho}}^{\rm wav}$, which can also cause the saturation of mass-loss rate.
The detailed parameter survey on turbulent correlation length should be done to test the above hypothesis.

We note that, because radiation dominates the energetics of the chromosphere, any dissipated energy is quickly radiated away.
In this sense, our model is consistent with \citet{Suzuk13}, with the saturation of mass-loss rate being compensated by an enhanced radiative loss.

To summarize our analysis, we have shown that a larger fraction of Alfv\'en wave energy is dissipated in the chromosphere for faster-rotating cases.
As a result, even though the photospheric Alfv\'en-wave luminosity follows a power-law relation in $\Omega_\ast$,
the coronal-base Alfv\'en-wave luminosity saturates with increasing rotation, thus producing a saturation in the mass-loss rate.

\section{Discussion} \label{sec:discussion}

\subsection{Comparison with Cranmer $\&$ Saar (2011)}
A standard theoretical model of the stellar-wind mass-loss rate, for low-mass stars, is given by \citet{Cranm11}.
Although both our model and \citet{Cranm11} are based on Alfv\'en wave heating and are calibrated by solar wind observations,
the rotation dependence of the mass-loss rate is different. For example, when $P_{\rm rot} = 2 {\rm \ day}$, 
the \citet{Cranm11} model yields a mass-loss rate that is 100 times larger than that predicted by our model.
There are three factors that explain this difference.
\begin{enumerate}
\item
\citet{Cranm11} assumed a steeper dependence of $f^{\rm open}_\ast$ on ${\rm Ro}$ with the exponent ranging between $-2.5$ and $-3.4$ in the unsaturated regime, while our model assumes much weaker dependence: $f^{\rm open}_\ast \propto {\rm Ro}^{-1.2}$.
For example, when ${\rm Ro} / {\rm Ro}_\odot = 0.1$, $f^{\rm open}_\ast = 0.36$ in \citet{Cranm11} while $f^{\rm open}_\ast = 0.016$ in this work.
Given that the mass-loss rate approximately scales as $\propto \left( f^{\rm open}_\ast \right)^{5/7}$, this discrepancy yields a factor of $9.3$ difference 
between \citet{Cranm11} and our model.
Since many observational aspects of stellar magnetism/winds are unresolved, theoretical mass-loss rates remains uncertain by around a factor of $10$.
\item
\citet{Cranm11} employed a simplified model of Alfv\'en wave propagation.
Although their model also considers the turbulent dissipation of Alfv\'en waves, the difference between their wave equations, and those used in this work, may lead to discrepancies in the resulting mass-loss rates.
This hypothesis should be tested in future by directly comparing the coronal wave energy between \citet{Cranm11} and our model.
\item
\citet{Cranm11} assumed that the wind speed is constant regardless of the open-flux filling factor.
In reality, even without rotational acceleration, the wind speed tends to be higher for larger open-flux filling factors, which is explained as follows.
Faster rotators exhibit larger coronal Alfv\'en speed that allows more heat deposited beyond the sonic point.
Given that the kinetic energy flux of the wind ($\propto \rho v_r^3$) is constant, the mass-loss rate ($\propto \rho v_r$) becomes smaller as the wind velocity $v_r$ increases in response to enhanced heating in the supersonic region.
\end{enumerate}

\subsection{Magnetic transient events}
Our model assumes that the global magnetic structure is invariant on timescales of the stellar wind acceleration.
However, actual stellar magnetic fields can evolve in comparable or even shorter timescales than the wind acceleration.
Specifically, the large scale shuffling of magnetic field lines by super-granular motions is observed to cause magnetic reconnection and open closed magnetic features \citep{Fisk099,Antio11,Moore11,Higgi17}.
Although the reconnection/loop-opening process is unlikely to be able to drive the majority of the solar wind \citep[see, e.g.,][]{Cranm10,Lione16}, 
it may play an important role for more active stars.
Additionally, the reconnection/loop-opening process can work indirectly.
For example, if the open magnetic field (carrying a quasi-steady wind) is rapidly connecting to closed loops with a high temperature, 
then the wind properties are determined by the closed loop temperature \citep{Fionn18}.
Such reconnection can also feed magnetohydrodynamic waves in addition to the surface granular motion \citep{Cranm18}.

Eruptive processes such as coronal mass ejections (CMEs) can also be important in active stars \citep{Aarni12,Drake13}.
According to \citet{Cranm17a}, CMEs could be a dominant source of mass loss for moderately faster rotators than the Sun.
For much younger, much faster rotators, the centrifugally-supported "slingshot prominence" are also expected to be present \citep{Colli89a,Colli89b}, 
and are likely to play a significant role in mass loss and magnetic braking.
Recently \citet{Jardi19} have extended the $F_X$-$\dot{M}_w$ relation of \citet{Wood014} to more active stars, 
based on mass-loss rates estimated from slingshot prominences. 
They show a significant mass-loss through prominence ejection for such rapid rotators.
The role of these eruptive processes should be taken into account in future works.

\subsection{Comparison with observations of spin evolution}
In spite of successfully reproducing the stellar spin down $\Omega_\ast \propto t^{-0.55}$,
there exist several discrepancies between our model and stellar observations of spin evolution.
As already mentioned in Section \ref{sec:result_torque}, 
the torque is smaller than the empirical value from stellar observation \citep{Matt015} but is consistent with solar wind observations \citep{Finle18b}.
Since our model is calibrated by the solar wind, the deviation of our model from \citet{Matt015} might be a result of the solar magnetic field having an unusual character.
Though reconstructions of the solar open magnetic flux from the last 9000 years also recover the same solar wind torque as our model \citep[see][]{Finle19b}.

Indeed, recent asteroseismic observations indicate that the solar dynamo could be in transition \citep{Metca16}.
The deviation between our model and \citet{Matt015} would be explained if the dynamo transition works to reduce the amount of open magnetic flux, 
by a factor of 2.4 from the canonical value.
This hypothesis is consistent with the observed break-down of gyrochronology \citep{Sader16}, 
and some spin-down models already take this effect into account \citep[e.g.][]{Garra18}.
However we must note that, from the perspective of dynamo simulations, large-scale field diminishing at ${\rm Ro}>1$ is not supported \citep{Strug18,Warne18,Guerr19}.
Thus, the small solar torque could be attributed to another mechanism.

\subsection{Implications for stars in the saturated-regime}

We have assumed that the filling factor of the open flux regions monotonically increases with rotation rate, Eq. (\ref{eq:fopen_thiswork}), which is derived from stars in the slow-rotator regime of \citet{See0019b}.
However, this power-law relation may be modified for fast rotators; 
the filling factor of the open regions may saturate at rapid rotation rates because a large fraction of the surface is expected to be covered by closed loops (these closed loops are thought to provide the observed coronal X-ray flux).
This modification would affect the magnetic field strength in the chromosphere and the low corona, 
and changes the vertical profile of the Alfv\'en velocity there.
A different profile of $v_{\rm A}$ may enhance the transmitted fraction of Alfv\'enic waves through the transition region \citet{Suzuk13}, 
which could increase the mass loss rate in the fast-rotator regime. 

In this work, we focused on the unsaturated regime of magnetic activity.
However, many young low-mass stars (especially M dwarfs) lie in the saturated activity regime \citep[e.g.][]{Wrigh16,See0019b}.
Given that the total open flux might be constant for stars in the saturated regime, our simulation results yield several implications for the winds of these stars. We expect the mass-loss rate should be constant in the saturated regime.
As shown in Figure \ref{fig:energetics_rot_dependence}, as long as the total open flux is fixed, increasing rotation rate does not yield larger mass-loss rates.
Instead, the increased rotational energy is used purely for wind acceleration.
When the wind velocity is larger, $v_{r,A}$ (wind velocity at Alfv\'en point) should also be larger.
According to the analytical expression of the torque in Eq. (\ref{eq:torque_scaling1}), a larger $v_{r,A}$ yields a smaller $\tau_w / \Omega_\ast$.
Thus in the saturated regime, the torque could have a weaker-than-linear dependence on $\Omega_\ast$.
In future, this prediction can be directly tested by numerical simulations of stars in the saturated-regime.

\vspace{1em}
Numerical computations were carried out on Cray XC50 and PC cluster at Center for Computational Astrophysics, National Astronomical Observatory of Japan.
MS is supported by Grant-in-Aid for Japan Society for the Promotion of Science (JSPS) Fellows and 
by the NINS program for cross-disciplinary study (Grant Nos. 01321802 and 01311904) 
on Turbulence, Transport, and Heating Dynamics in Laboratory and Solar/Astrophysical Plasmas: "SoLaBo-X".
TKS is supported in part by Grants-in-Aid for Scientific Research from the MEXT of Japan, 17H01105.
SPM, VS, and AJF are supported by the European Research Council, under the European Union's Horizon 2020 research and innovation program (agreement No. 682393, AWESoMeStars).
AAV is supported by the European Research Council, under the European Union's Horizon 2020 research and innovation program (agreement No. 817540, ASTROFLOW).
AS and ASB acknowledge funding by ERC WHOLESUN 810218 grant, INSU/PNST, CNES-PLATO and CNES Solar Orbiter. 
AS acknowledges funding from the Programme National de Planétologie (PNP).
VR is funded by the ERC SLOW{\_}\,SOURCE project (SLOW{\_}\,SOURCE - DLV-819189).
This work benefitted from discussions within the international team "The Solar and Stellar Wind Connection: Heating processes and angular momentum loss", supported by the International Space Science Institute (ISSI).

\begin{appendix}
\section{Derivation of basic equations}\label{app:derivation}
Except the non-ideal terms (gravity, radiative loss, thermal conduction, turbulent dissipation), our basic equations Eq.s (\ref{eq:basic_ro})-(\ref{eq:basic_en}) are derived from the typical ideal MHD equations as follows.
Given the metrics of
\begin{align}
    h_r = 1, \ \ \ \ h_\theta = rf^{\rm open}, \ \ \ \ h_\phi = r, \label{eq:appendix:metrics}
\end{align}
and considering a one-dimensional system ($\partial / \partial \theta = \partial / \partial \phi = 0$), the nabla operators are expressed as follows.
\begin{subequations}
\begin{align}
    \nabla \psi &= \frac{\partial \psi}{\partial r} \vec{e}_r, \label{eq:nabla_grad}  \\
    \nabla \cdot \vec{A} &= \frac{1}{r^2 f} \frac{\partial}{\partial r} \left( r^2 f A_r \right), \label{eq:nabla_div}  \\
    \nabla \times \vec{A} &=  
    - \vec{e}_\theta \frac{1}{r}  \frac{\partial}{\partial r} \left( r A_\phi \right) 
    + \vec{e}_\phi \frac{1}{r f} \frac{\partial}{\partial r} \left( r f A_\theta \right), \label{eq:nabla_rot}
\end{align}
\end{subequations}
where, for simplicity, we denote $f^{\rm open}$ as $f$.
Using these expressions, each basic equation is derived in a straightforward manner.
For example, the inertial and Lorentz forces in the equation of motion are written explicitly as
\begin{align}
	\left( \rho \vec{v} \cdot \nabla \right)  \vec{v} &= \rho \left( \nabla \times \vec{v} \right) \times \vec{v} + \frac{1}{2} \rho \nabla \left( v^2 \right) \nonumber \\
	&= \vec{e}_r \left[ \rho v_r \frac{\partial}{\partial r} v_r - \rho v_\theta^2 \frac{d}{dr} \ln \left( r f \right) - \rho v_\phi^2/r\right] +  \vec{e}_\theta \frac{\rho v_r}{r f} \frac{\partial}{\partial r} \left( r f v_\theta \right) + \vec{e}_\phi \frac{\rho v_r}{r} \frac{\partial}{\partial r} \left( r v_\phi \right), \\
        \frac{1}{4\pi} \left(\nabla \times \vec{B} \right) \times \vec{B} &= \vec{e}_r \left[ - \frac{B_\theta}{4 \pi r f} \frac{\partial}{\partial r} \left( B_\theta r f \right) - \frac{B_\phi}{4 \pi r} \frac{\partial}{\partial r} \left( r B_\phi \right) \right] + \vec{e}_\theta \frac{B_r}{4 \pi rf} \frac{\partial}{\partial r} \left( r f B_\theta \right) + \vec{e}_\phi \frac{B_r}{4 \pi r} \frac{\partial}{\partial r} \left( r B_\phi \right).
\end{align}
Similarly, the rotation of the electromotive force is
\begin{align}
	\nabla \times \left( \vec{v} \times \vec{B} \right) = 
	- \vec{e}_\theta \frac{1}{r} \frac{\partial}{\partial r} \left[ r \left(v_r B_\theta - v_\theta B_r \right) \right]
	- \vec{e}_\phi \frac{1}{rf} \frac{\partial}{\partial r} \left[ rf \left(v_r B_\phi - v_\phi B_r \right) \right], 
\end{align}
After these calculations, one can obtain the basic equations after rewriting in a conservation form.

\vspace{1em}
\section{An analytical formulation of coronal rotational luminosity } \label{app:rotational_luminosity}
The coronal rotational luminosity $L_{A,{\rm cor}}^{\rm rot}$ is obtained analytically based on Weber-Davis solution.
We begin with the analytical expression of (time-averaged) $v_\phi$ and $B_\phi$:
\begin{align}
	v_\phi = r \Omega_\ast \frac{M_A^2 L / (r^2 \Omega_\ast) -1}{M_A^2 -1}, \ \ \ \ B_\phi = \frac{B_r}{v_r} \left( v_\phi - r \Omega_\ast \right).
\end{align}
Near the coronal base where the wind velocity is negligibly small, we can approximate $v_\phi$ and $B_\phi$ to the first order of $M_A^2$ as
\begin{align}
	v_\phi \approx r \Omega_\ast \left( 1 - \frac{M_A^2 r_A^2}{r^2} \right), \ \ \ \ B_\phi \approx -r \Omega_\ast \frac{B_r M_A^2 r_A^2}{v_r r^2}.
\end{align}
To the first order of $M_A^2$, the rotational luminosity at the coronal base is given as
\begin{align}
	L_{A,{\rm cor}}^{\rm rot} = - \left. 4 \pi r^2 f^{\rm open} \cdot \frac{B_r}{4 \pi} v_\phi B_\phi \right|_{\rm cor} \approx r_A^2 \Omega_\ast \dot{M}_w = \tau_w \Omega_\ast.
\end{align}
It is interesting to see that the rotational luminosity is approximated by $\tau_w \Omega_\ast$.
Using Eq. (\ref{eq:identity_rA2Mdot}), the above formulation is further simplified as
\begin{align}
	L_{A,{\rm cor}}^{\rm rot} \approx \frac{\Phi_{\rm open}^2}{16 \pi^2 v_{r,A}} \Omega_\ast^2 \propto \Omega_\ast^{3.83},
\end{align}
where we have used $\Phi_{\rm open} \propto \Omega_\ast^{1.2}$ and $v_{r,A} \propto \Omega_\ast^{0.57}$.

\end{appendix}

\bibliographystyle{apj}

\end{document}